\documentclass[a4paper,fleqn]{cas-dc}

\usepackage[numbers]{natbib}

\usepackage{booktabs}
\usepackage{tikz}
\usepackage{kotex}

\usepackage{multirow}
\usepackage{amssymb}
\usepackage{pifont}
\usepackage{array}
\usepackage{wasysym}
\usepackage{color}
\usepackage{tabularx}
\usepackage{subfig}
\usepackage{soul}
\usepackage{array} %

\newcolumntype{L}[1]{>{\raggedright\arraybackslash}p{#1}}

\newcommand{\ourtool}{\texttt{MetaScanner}}
\newcommand*\circled[1]{\tikz[baseline=(char.base)]{
    \node[shape=circle,draw,inner sep=0.5pt] (char) {#1};}}

\newcolumntype{M}[1]{>{\centering\arraybackslash}m{#1}}

\usepackage{listings}
\usepackage{xcolor}
\usepackage{caption}
\lstdefinelanguage{ReportLog}{
    basicstyle=\ttfamily\footnotesize,
    breaklines=true,
    frame=single,
    rulecolor=\color{black!30},
    backgroundcolor=\color{gray!5},
    keywordstyle=\color{red!70!black}\bfseries,
    morekeywords={DETECTED, WARNING, MALICIOUS, SUSPICIOUS, MISMATCH},
    moredelim=**[is][\color{blue!60!black}]{@@}{@@}, %
    moredelim=**[is][\color{green!40!black}]{!}{!}, %
}%

\def\tsc#1{\csdef{#1}{\textsc{\lowercase{#1}}\xspace}}
\tsc{WGM}
\tsc{QE}

\begin{document}
\sloppy
\setlength{\emergencystretch}{3em}

\let\WriteBookmarks\relax
\def\floatpagepagefraction{1}
\def\textpagefraction{.001}

\shorttitle{}    

\shortauthors{Lee et al.}  

\title [mode = title]{MetaScanner: An Empirical Study and Proactive Defense against Deceptive UI Attacks in Social VR}

\tnotemark[1] 

\tnotetext[1]{This work extends a poster entitled "Illusion Worlds: Deceptive UI Attacks in Social VR~\cite{lee2025illusion}" presented at the \textit{2025 IEEE Conference on Virtual Reality and 3D User Interfaces Abstracts and Workshops (VRW)}.}

\credit{}

\affiliation[cbnu]{organization={Chungbuk National University},
addressline={1 Chungdae-ro, Seowon-gu}, 
city={Cheongju, Chungcheongbuk-do},
postcode={28644}, 
country={Republic of Korea}}

\affiliation[kwu]{organization={Kwangwoon University},
            addressline={27-15, Gwangun-ro}, 
            city={Nowon-gu},
            postcode={01897}, 
            state={Seoul},
            country={Republic of Korea}}

\affiliation[etri]{organization={ETRI},
            addressline={218, Gajeong-ro}, 
            city={Yuseong-gu},
            postcode={34129}, 
            state={Daejeon},
            country={Republic of Korea}}

\author[cbnu]{Junhee Lee}[
        orcid=0009-0009-4967-3621
]
\ead{junhee1004m@gmail.com}

\author[etri]{Seungwon Woo}[
        orcid=0009-0001-8874-6936
]
\ead{seungww@etri.re.kr}
\credit{}

\author[kwu]{Minseok Kim}[
        orcid=0009-0004-3991-6486
]
\ead{rhdtls1000@kw.ac.kr}
\credit{}

\author[kwu]{Jongseop Kim}[
        orcid=0009-0000-5589-3588
]
\ead{joseab5965@kw.ac.kr}
\credit{}

\author[etri]{Kiyoung Kim}[
        orcid=0000-0002-4372-0943
]
\ead{kykim@etri.re.kr}
\credit{}

\author[etri]{Hwanjo Heo}[
        orcid=0000-0002-8105-4224
]
\ead{hwanjo@etri.re.kr}
\cormark[1]
\credit{}

\author[cbnu]{Jinwoo Kim}[orcid=0000-0003-1303-8668]
\cormark[1]
\ead{jinwookim@cbnu.ac.kr}
\credit{}

\cortext[1]{Co-corresponding authors}

\begin{abstract}
Social virtual reality (VR) platforms have surged in popularity but continue to face significant security challenges, including privacy invasion and user manipulation. Yet, the security risks posed by deceptive content remain largely unexplored. In this paper, we introduce four novel deceptive UI attacks that covertly manipulate social VR users into performing unintended actions without their awareness. We show how adversarial content creators can embed seemingly benign scripts and interactive elements within carefully designed virtual worlds, deceiving users into unknowingly carrying out harmful behaviors. We implemented these attacks on VRChat, a leading social VR platform, and an IRB-approved user study involving 30 participants confirmed their effectiveness. To mitigate these threats, we propose \ourtool{}, a proactive analysis system that efficiently scans virtual world assets, including scripts and interactive objects, to identify suspicious or deceptive elements before users are exposed to them. We demonstrate that \ourtool{} can successfully detect all threats in test worlds where the four attack scenarios were implemented. Additionally, \ourtool{} shows its lightweight nature, with an average scanning time of 5 seconds across 38 real-world assets.

\end{abstract}


\maketitle

\section{Introduction}

Over the past decade, social virtual reality (VR) platforms have grown rapidly in popularity, driven by the immersive experiences enabled by head-mounted displays (HMDs). These devices allow users to explore and interact within user-generated virtual environments that support rich social interactions. In these spaces, users communicate through customizable 3D avatars with body tracking, fostering personal expression and a stronger sense of presence. Among these platforms, VRChat~\cite{vrchat} has become the most prominent, reaching over 40,000 peak users in recent years~\cite{vrchat_user,vrchat_stats}.

However, the success of VR has introduced unprecedented security concerns. The research community has reported various threats that exploit user interactions in VR applications. For example, previous studies have demonstrated privacy breaches and device exploitation, such as inferring personal information from head-tracking data~\cite{slocum2023going}, performance metrics~\cite{zhang2023head}, and vital signs~\cite{zhang2023facereader}. Given the close physical proximity of VR devices to the human body, security risks in VR environments are particularly critical.

\begin{figure}
    \centering
    \includegraphics[width=\linewidth]{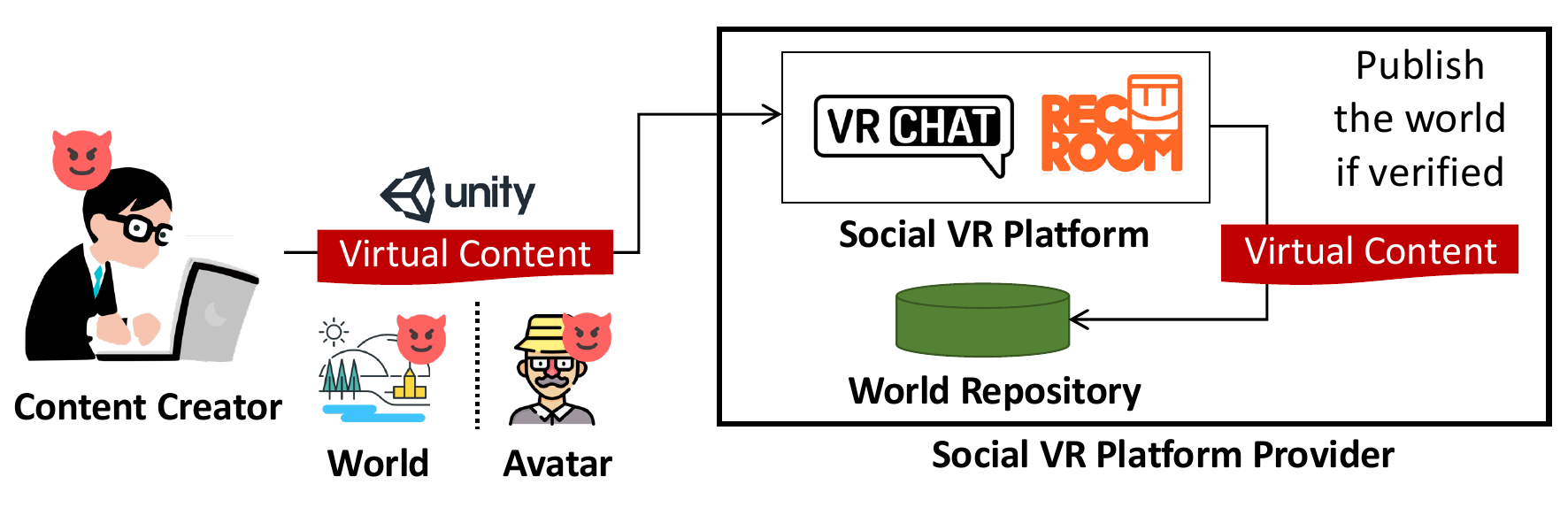}
    \caption{Overview of our threat model---a social VR ecosystem in which a \emph{content creator} (acting as an adversary) generates and uploads malicious virtual content.}
    \label{fig:threat_model}
\end{figure}

While these privacy-related threats in VR have received considerable attention, little is known about security risks arising from \emph{deceptive user interface (UI)}. The risks of deceptive UI have recently been discussed in the VR community, particularly in the context of intentionally manipulating user decisions to maximize corporate profit~\cite{hadan2024survey}. Existing studies have shown that deceptive design can mislead users into making incorrect decisions~\cite{cheng2023exploring,lee2021adcube,cheng2024user}, or even cause physical and psychological harm~\cite{tseng2022dark,yang2024inception} by distorting their perception of virtual environments or user interfaces.  However, its impact on social VR platforms remains underexplored, even though these platforms rely on user-generated content—creating opportunities for adversaries to embed deceptive UI elements more easily.\looseness=-1

In this paper, we examine a fundamental characteristic of social VR platforms that makes deceptive UI attacks particularly practical and easy to deploy: users can freely create virtual content, potentially with malicious intent (see Figure~\ref{fig:threat_model}). Specifically, an \emph{adversarial content creator} may exploit three core components of the social VR environment: (i) virtual worlds, (ii) avatars, and (iii) user interactions with these elements. These components can embed harmful logic while appearing benign, leveraging deceptive features such as game-like settings, visually indistinguishable objects, and hidden scripts—augmented by social engineering tactics that mask malicious intent. As users are encouraged to explore a wide array of virtual worlds to satisfy their social needs~\cite{sykownik2021most}, such attacks can effectively lure and engage unsuspecting participants.

Specifically, we found that an adversarial content creator can manipulate visual and spatial properties of user-generated content, such as by placing invisible or overlapping objects, to intercept, suppress, redirect, or replace users' expected interactions and exploit their visual trust. To this end, we propose four novel UI attacks targeting these properties in social VR platforms: (i)~\emph{Layered Click Redirection}, where an adversary’s object captures user inputs and redirects them to unintended content (e.g., the adversary's ads or videos); (ii)~\emph{Raycasting Phishing}, where an invisible object obstructs user interactions to deceive users; (iii)~\emph{In-World Input Hijack}, where an adversary’s object covertly exfiltrates sensitive information (e.g., credit card numbers or passwords) to a hidden room within the virtual world; and (iv)~\emph{Avatar Quishing}, where an adversary manipulates their avatar to entice users into interacting with malicious QR codes embedded in the avatar. We design and implement these attacks by leveraging the unique characteristics of social VR. 

Unlike traditional UI attacks in web or standalone VR that primarily rely on visual obfuscation~\cite{lee2021adcube,cheng2024user}, our threat model exploits the unique vulnerability of social immersion and active interaction. In social VR, users are not merely observers but active participants who psychologically inhabit the virtual space~\cite{li2023interactions,kou2023harmful,dwivedi2022metaverse}. This cognitive load from social interactions (e.g., chatting, gaming) significantly lowers users' vigilance against deceptive UI. Consequently, our proposed attacks leverage this implicit trust by leading users to perceive a fake system warning (e.g., Raycasting Phishing) or a malicious avatar's QR code (e.g., Avatar Quishing) as legitimate environmental cues rather than external threats.

We conducted an IRB-approved user study with 30 participants, of whom 16 had prior VR experience and 14 had never used a VR headset. Their feedback shows that our attacks effectively deceived participants regardless of their prior VR experience. As VR adoption grows and attracts more users with lower digital literacy, who are known to be more susceptible to scams~\cite{mouncey2025phishing,graham2017capable}, these results underscore the need for user education to mitigate risks from deceptive UI attacks. All four attacks were deployed in a proof-of-concept virtual world on VRChat. To the best of our knowledge, this is the first work to exploit the social VR ecosystem where users can act as content creators with potentially malicious intent.

To mitigate these risks, we introduce a proactive and lightweight countermeasure called \ourtool{}. \ourtool{} is designed to rapidly detect suspicious objects—such as those containing irrelevant scripts, invisible elements, or malicious URLs—within a virtual world and provide visual alerts to help users avoid potentially harmful consequences. We evaluated the defense capabilities of \ourtool{} by demonstrating that it successfully detected all four proposed attack scenarios. Additionally, we validated the scanning performance of \ourtool{} on 38 collected virtual worlds, showing an average processing time of approximately 5 seconds. This article extends our preliminary poster~\cite{lee2025illusion}.

\noindent\textbf{Contributions.} Our contributions are summarized as follows:  
\begin{itemize}
    \item We introduce four novel UI attacks that can be covertly deployed by an adversarial content creator (Section~\ref{section:ui_attacks}). 

    \item We demonstrate the feasibility of the proposed attacks by implementing a custom virtual world on VRChat, the most widely used social VR platform. Also, we validate the effectiveness of our attacks through an IRB-approved user study involving 30 participants, of whom 16 had prior VR experience and 14 had never used a VR headset (Section~\ref{section:user_study}).

    \item We propose an effective and lightweight defense system \ourtool{}\footnote{\ourtool{} is publicly available at \url{https://github.com/cnslab-cbnu/MetaScanner}. The version used in this paper is archived at \url{https://doi.org/10.5281/zenodo.24151657}.} that proactively detects prospective UI attacks before adversarial content goes online (Section~\ref{sec:metascanner}).
\end{itemize} %
\section{Background}

In this section, we provide the necessary background to contextualize our attacks and introduce several key terms relevant to our study.

\subsection{Social VR Platforms}

Social VR platforms are widely used applications that offer users immersive 3D experiences through modern head-mounted displays (HMDs). These platforms typically consist of three core components. First, users are represented by \emph{avatars}, which serve as a medium for self-expression while maintaining anonymity. Avatars can be customized with characters, clothing, accessories, and other creative designs, and are often shared and distributed within the platform's ecosystem. Second, users can explore \emph{virtual worlds} designed around diverse themes—such as games, markets, or social spaces—populated with interactive virtual objects. Similar to avatars, users are also empowered to create and upload their own virtual worlds. Third, users engage in \emph{interactions} within these environments, including communicating with others (i.e., avatars), viewing media content, and triggering in-world events through actions involving VR controllers or hand gestures.

VRChat~\cite{vrchat}—the most popular social VR platform to date—exemplifies this model by supporting extensive user-generated content and immersive social interaction. Users can freely explore virtual worlds created by others while embodying personalized avatars. VRChat also provides a Software Development Kit (SDK) that enables users to develop and upload their own avatars and virtual worlds. Furthermore, VRChat avatars support rich expressiveness through features such as lip-syncing, eye tracking, and blinking, thereby enhancing users' sense of presence and social connection within the virtual environment.

\subsection{Security Concerns in VR}

VR has recently garnered significant attention due to its unique security and privacy challenges. The tight integration of VR devices—such as head-mounted displays (HMDs) and handheld controllers—with the user creates new avenues for adversaries to compromise user safety and extract personal information. For instance, attackers can exploit motion sensor data from HMDs to infer head movements, facilitating attacks such as motion-based keylogging~\cite{slocum2023going} or the extraction of sensitive information~\cite{zhang2023head}. 

Recent studies have also demonstrated that physiological signals, including vital signs, can be mined from VR systems~\cite{zhang2023facereader}, further exacerbating privacy concerns.  
Beyond data leakage, the immersive nature of VR introduces new vectors for psychological and physical harm. Perception-based attacks that manipulate the user’s field of view—by inserting misleading or unexpected virtual objects—can result in physical injuries~\cite{tseng2022dark} or provoke critical decision-making errors~\cite{cheng2023exploring}. Moreover, adversaries can repurpose virtual objects for ad fraud, deceiving users into interacting with objects that mimic legitimate advertisements~\cite{lee2021adcube}.

We posit that similar, yet more potent, attacks leveraging \emph{deceptive UI} can be executed within social VR platforms. With their large user bases and emphasis on user-generated content, social VR platforms offer an attractive and scalable attack surface for adversaries seeking to exploit these vulnerabilities.

\section{Threat Model and Assumptions}
\label{subsec:motivation}

\noindent\textbf{Threat Model.} We assume an adversary whose objective is to carry out malicious activities on popular social VR platforms (e.g., VRChat or Rec Room) by creating and distributing adversary-generated virtual content. User-generated content (UGC) in social VR fosters implicit trust, shapes community norms and user behavior~\cite{kou2023harmful,dwivedi2022metaverse}, and—by sustaining a high degree of immersion—reduces users’ awareness of security risks~\cite{li2023interactions}, making them more susceptible to manipulation. Together, these properties enable a range of attacks (social manipulation, deceptive UI elements, data exfiltration, etc.) that exploit users’ lowered risk perception and trust in UGC. Given the large user base, such an adversary could deceive users into engaging in illegal activities or extract sensitive personal information; our threat model, therefore, considers a remote adversary who accesses the platform like any legitimate user, while server-side compromises or attacks on platform infrastructure are beyond the scope of this work.

\vspace{0.05in}

\noindent\textbf{Attack Deployment.}
An adversary can craft virtual content embedded with malicious logic and upload it using the platform's Software Development Kit (SDK), which enables the creation of avatars, objects, and virtual worlds. Depending on the attack, the adversary may control either an entire virtual world or only adversary-crafted content, such as an avatar, within an existing world. The social VR ecosystem makes this particularly effective, as adversary-crafted content can persist in the environment and trigger attacks covertly. Figure~\ref{fig:threat_model} illustrates this workflow, in which the adversary, acting as a content creator, develops virtual content containing malicious objects and scripts and uploads it to the platform.
The content undergoes platform verification, focusing mainly on overtly harmful elements (e.g., sexual or hate-related content). As part of this process, all uploaded virtual worlds are reviewed to ensure compliance with platform policies and to prevent the distribution of harmful content (e.g.,~\cite{vrchat_community_guideline,vrchat_creator_guideline}). The verification typically focuses on identifying explicitly harmful elements such as sexual or hate-related content. If approved, it is published for user access. Covertly malicious objects may bypass this process, and while users can report suspicious content, the execution of attacks or exposure of sensitive information without user awareness remains a critical concern. %
\section{Attack Design}
\label{section:ui_attacks}

In this section, we introduce four UI attack techniques that can be employed by adversary-crafted content in social VR platforms. 
The attacks were derived through a three-part methodology: (i) a systematic decomposition of the social-VR attack surface into its trust-bearing components of virtual worlds, avatars, and user interactions (Section~\ref{subsec:motivation}); (ii) adaptation of already established deception primitives from the web/AR/VR literature (e.g., clickjacking~\cite{huang2012clickjacking}, WebXR ad fraud~\cite{lee2021adcube}, and physical-world quishing); and (iii) hands-on analysis of the actual VRChat/Unity SDK.
Our analysis shows that all four attacks share a common characteristic: an adversary can manipulate visual and spatial properties of user-generated content, such as by placing invisible or overlapping objects, to intercept, suppress, redirect, or replace users' expected interactions and exploit their visual trust. The security risks of such user-generated content are not merely hypothetical: social VR platforms have already encountered malicious content such as crasher avatars~\cite{vrchat_creator_guideline, vrchat_community_guideline} that overload colliders to disrupt or eject nearby users, prompting the introduction of avatar trust and safety ranks~\cite{vrchat_safety,vrchat_eac}.

\subsection{Layered Click Redirection}
\label{subsec:clickjacking_design}

\begin{figure}
    \centering
    \includegraphics[width=\linewidth]{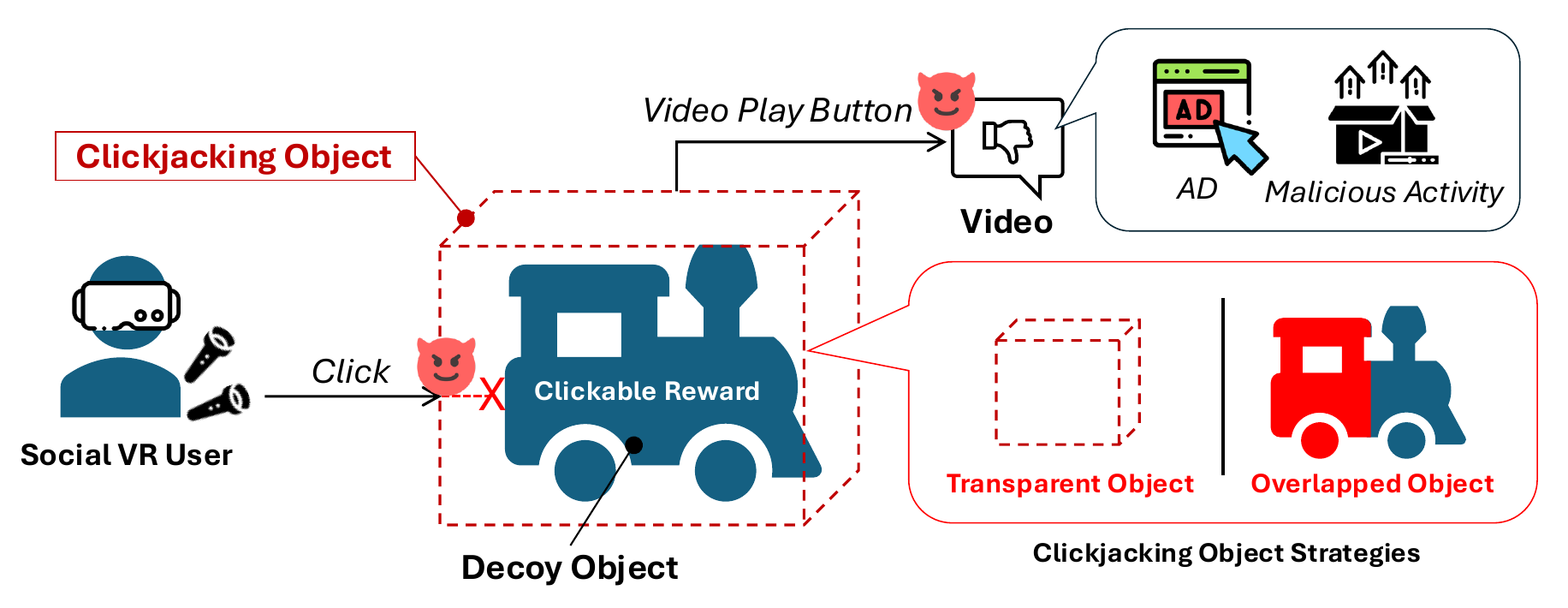}
    \caption{A layered click redirection in which the \emph{clickjacking object} intercepts user input intended for the \emph{decoy object}, redirecting it to trigger a video playback (or ads) without the user's awareness.}
    \label{fig:attack1}
\end{figure}

\begin{figure}
    \centering
    \subfloat[Before attack]{    
        \includegraphics[width=0.5\linewidth]{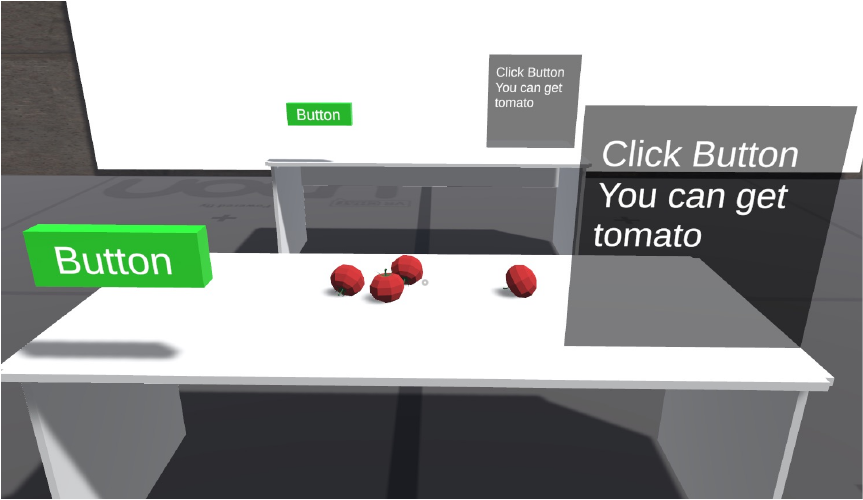}
        \label{fig:click_impl1}
    }
    \subfloat[After attack]{
    \includegraphics[width=0.5\linewidth]{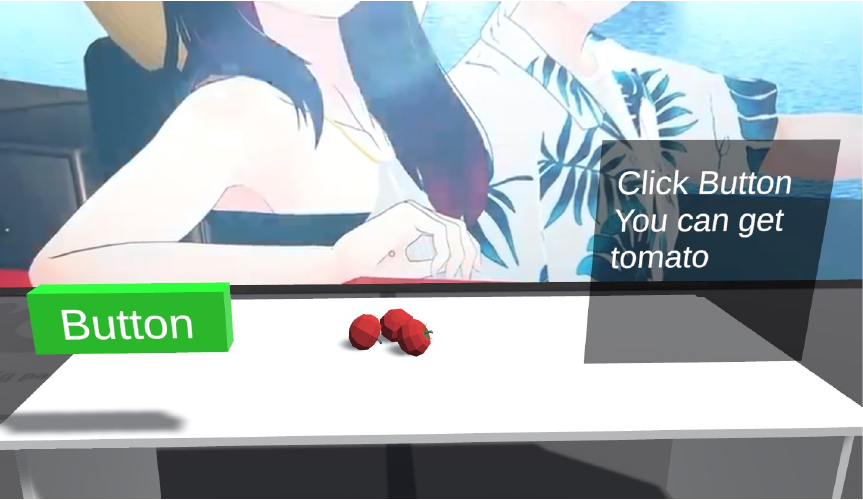}
        \label{fig:click_impl2}
    }
    \caption{Layered click redirection implementation.}
    \label{fig:click_impl}
\end{figure}

Clickjacking~\cite{huang2012clickjacking} is a well-established web-based attack technique in which users are deceived into clicking on seemingly harmless elements, while their actual interaction is hijacked by a transparent overlay (e.g., a \texttt{<div>} element). We observe that a more deceptive and impactful variant of this attack, which we call \emph{Layered Click Redirection}, can be executed within social VR environments. The immersive nature of VR leads users to become deeply engaged with virtual objects, rendering them especially vulnerable to visual deception. Figure~\ref{fig:attack1} illustrates a representative scenario, and Figure~\ref{fig:click_impl} shows how it appears in our VRChat implementation. In this scenario, an adversary places a carefully crafted \emph{decoy object} within a virtual world to attract user interaction. For instance, this object may appear as part of a mini-game, a clickable reward, or localized content (e.g., ads tailored by language or region). A \emph{clickjacking object} is then strategically positioned to intercept the user’s input.

To position the clickjacking object effectively, an adversary may adopt one of several placement strategies: (i)~\emph{a (shared) single object}, (ii)~\emph{a transparent object}, and (iii)~\emph{an overlapped object}. The first strategy is a naive method where an adversary implements both benign and malicious logic (i.e., clickjacking) within a single object. While this method is simple to deploy, it typically introduces a perceptible delay as the object processes dual logic, thereby increasing the likelihood of user detection. In contrast, the second strategy mirrors traditional web-based clickjacking techniques by placing a transparent object atop the decoy object. This approach benefits from modularity, as the transparent object can remain consistent across different decoys, enhancing the generalizability of the attack. If the decoy object risks detection---such as when user movement is obstructed or raycasting reveals inconsistencies---an adversary may instead employ the third strategy.

The last strategy involves rendering an object of identical size and position over the decoy, exploiting discrepancies between rendering and interaction orders across various rendering engines~\cite{cheng2024user}. In such cases, although the user visually perceives the decoy, their input is captured by the clickjacking object, making the deception significantly harder to detect. 
When the clickjacking object receives the user's input, it triggers an embedded event—such as launching a video or playing an advertisement—without the user's intent or awareness~\cite{lee2021adcube}. In crowded or content-rich virtual worlds, identifying the source of such behavior is particularly challenging, making the attack highly stealthy and effective. Furthermore, the overlapped object strategy introduces interaction inconsistencies depending on how each user engages with the object. For example, variations in raycasting angles or input modalities may cause one user to interact with the decoy object, while another inadvertently triggers the clickjacking object. These inconsistencies can, however, be controlled by the attacker through the object’s network synchronization mode. In a synchronized network mode, if a user interacts with the object, it becomes temporarily unavailable to others. By adjusting the script execution duration, the attacker can control how long the object remains unavailable, thereby regulating the number of users who can interact with it and be exposed to the attack at any given time. In contrast, in an asynchronous network mode, nearly all users can be exposed simultaneously, resulting in stable attack performance.

\vspace{0.05in}
\noindent\textbf{Attack Novelty.}
Unlike traditional clickjacking or similar attacks in standalone VR~\cite{cheng2024user}, our layered click redirection takes advantage of the implicit trust users place in user-generated content, which is a characteristic unique to social VR ecosystems. Notably, users are often guided or even compelled to perform specific actions by the framing or scenarios created within social VR virtual worlds. Moreover, by using an overlapped object strategy, the attack can selectively affect users based on their individual perspectives and interactions. This user-specific variability, combined with the distinctive features of social VR, leads to inconsistent experiences across users, making detection more difficult and hindering collective reporting.

\subsection{Raycasting Phishing}
\label{subsec:raycasting}

\begin{figure}
    \centering
    \includegraphics[width=\linewidth]{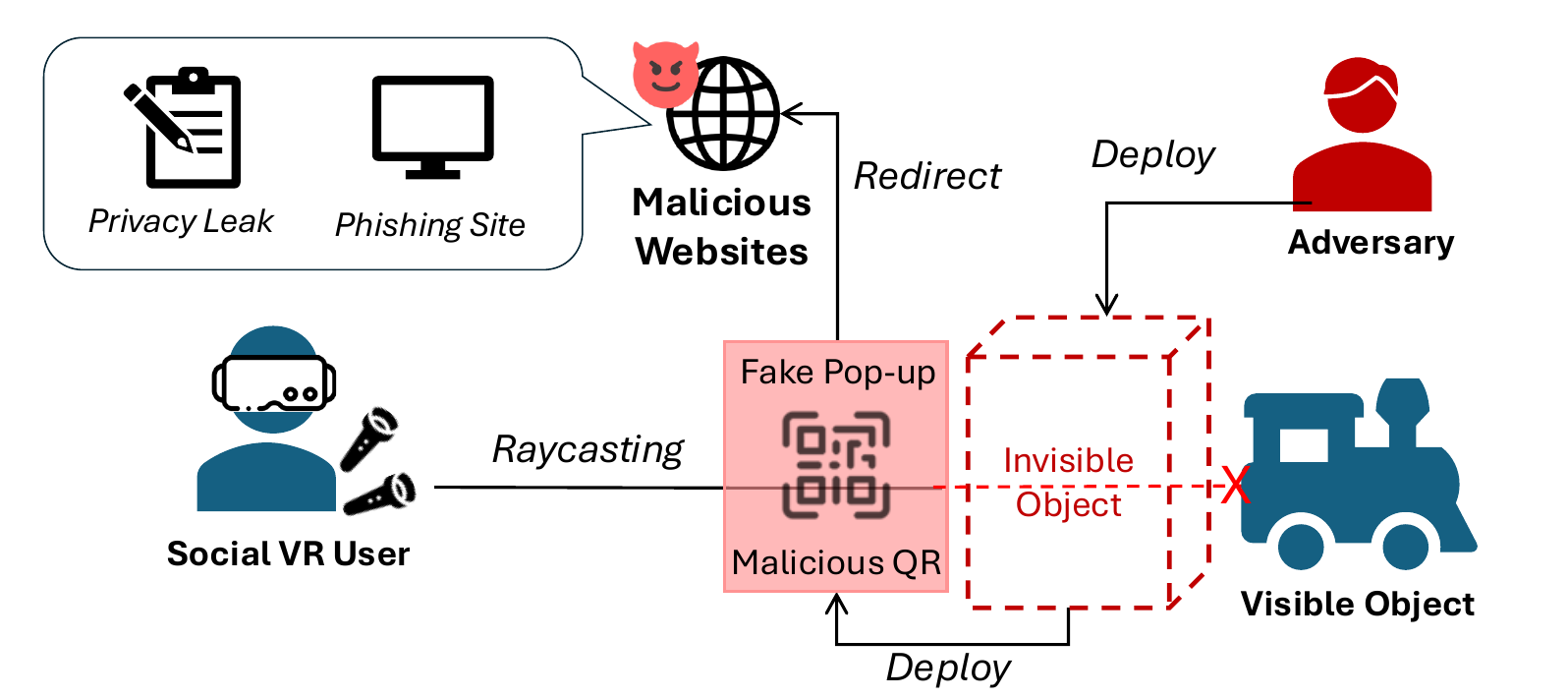}
    \caption{A raycasting phishing attack in which the adversary places an \emph{invisible object} to block the user's raycasting path, preventing interaction with \emph{visible objects}. Simultaneously, a \emph{fake UI} is displayed to mislead the user into performing a malicious action.}
    \label{fig:attack2}
\end{figure}

\begin{figure}
    \centering
    \subfloat[Before attack: the blue box is clickable.]{
        \centering
        \includegraphics[width=0.45\linewidth]{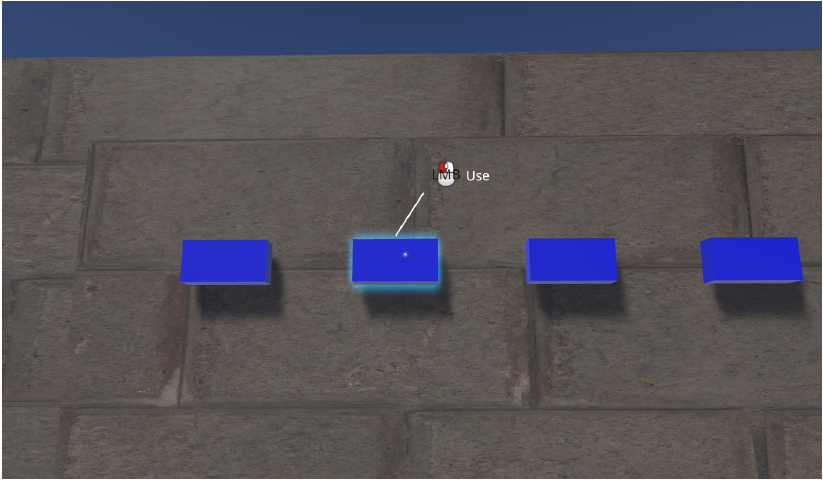}
        \label{fig:dos_imp1}
    }
    \hspace{0.05in}
    \subfloat[The trigger object (red box) for the attack.]{
        \centering
        \includegraphics[width=0.45\linewidth]{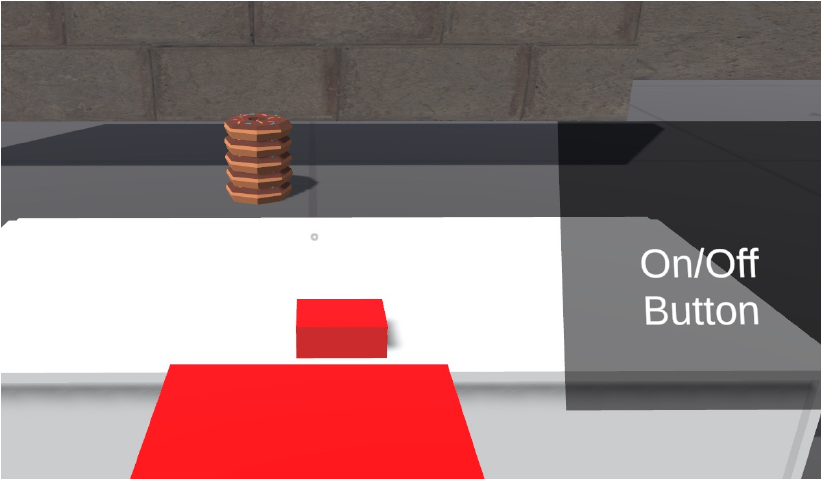}
        \label{fig:dos_impl2}
    }
    \vspace{-0.1in}
    \subfloat[After attack: the blue box is not clickable.]{
        \centering
        \includegraphics[width=0.45\linewidth]{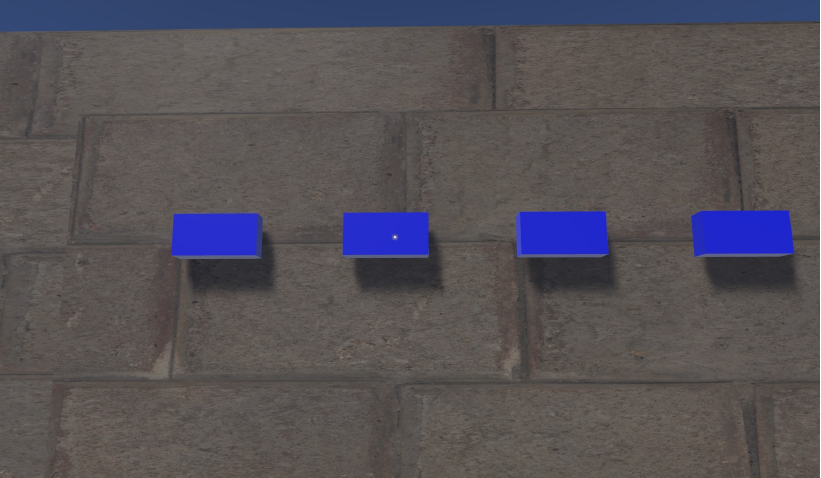}
        \label{fig:dos_impl3}
    }
    \hspace{0.05in}
    \subfloat[A banner displaying fake error info alongside a QR code.]{
        \centering
        \includegraphics[width=0.45\linewidth]{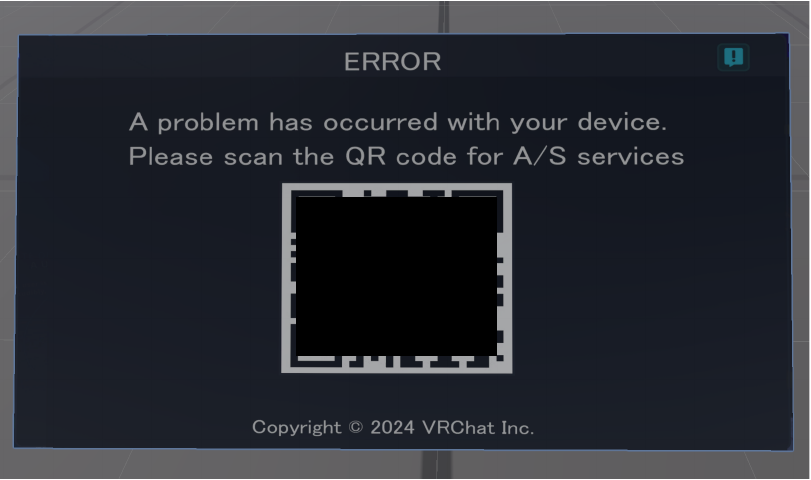}
        \label{fig:dos_impl4}
    }
    \caption{Raycasting phishing implementation.}
    \label{fig:dos_impl}
\end{figure}

In VR, users typically interact with objects using \emph{raycasting}, a technique that enables distant target selection. Virtual rays are generated through VR controllers or hand gestures, allowing users to point toward and interact with target objects—an essential mechanism for navigating and engaging with virtual environments. However, an adversary can exploit this mechanism by inserting an invisible object along the ray’s path, effectively blocking the user’s raycasting. This interference can prevent interaction with visible objects, creating an opportunity for a social engineering attack. By inducing failed interactions, the adversary may successfully mislead the user into believing that their controller is malfunctioning.

Figure~\ref{fig:attack2} illustrates a scenario involving a raycasting phishing attack, and Figure~\ref{fig:dos_impl} shows the corresponding scenes captured in our VRChat implementation. In this case, an invisible object is attached to the user, obstructing their raycasting and thereby preventing interaction with other visible objects in the environment. This obstruction is likely to confuse the user, as the invisible object offers no visual indication of interference—leading them to attribute the issue to a software or hardware malfunction. Exploiting this confusion, the adversary strategically places a deceptive notice nearby, containing a (fake) troubleshooting guide. The notice includes a QR code\footnote{QR codes are commonly placed on avatars and in-world objects in social VR platforms such as VRChat~\cite{vrchat-feedback-qr}.} that directs the user to a phishing website designed to resemble a legitimate platform provider or world creator. Consequently, the user may be tricked into clicking malicious links or following instructions that result in the installation of malware.

The adversary can adopt two primary strategies for deploying the invisible object: (i) attaching it to a user upon the fulfillment of a specific condition---such as completing a task in a game or through random selection---or (ii) attaching it by default when a user joins the virtual world. The first approach maintains stealth by selectively targeting a small number of users, thereby minimizing the likelihood of detection. In contrast, the second approach is designed for broader impact, aiming to affect as many users as possible. Furthermore, the adversary can manipulate the size of the invisible object to achieve different goals. It may be configured to cover either (i) the user’s entire avatar or (ii) a specific body part, such as one of the avatar’s hands. In the latter case, the user is more likely to perceive the issue as a malfunction of a single controller, reinforcing the deceptive nature of the attack.

\vspace{0.05in}
\noindent\textbf{Attack Novelty.}
While similar attacks that block users’ raycasting have been explored in previous studies~\cite{cheng2024user,sajid2025just}, 
our approach is novel in how it extends the deception beyond the initial disruption. 
Specifically, it employs a pop-up banner to effectively guide users into performing targeted actions, 
leveraging the immersive trust unique to social VR environments. 
Furthermore, unlike traditional web-based scareware or fake alert tactics, which rely on scripted pop-ups in flat interfaces, 
our attack exploits spatial interactions and user-generated content in social VR, making the deception 
even more convincing and harder to detect collectively.

\subsection{In-World Input Hijack}
\label{subsec:oitm}

\begin{figure}
    \centering
    \includegraphics[width=\linewidth]{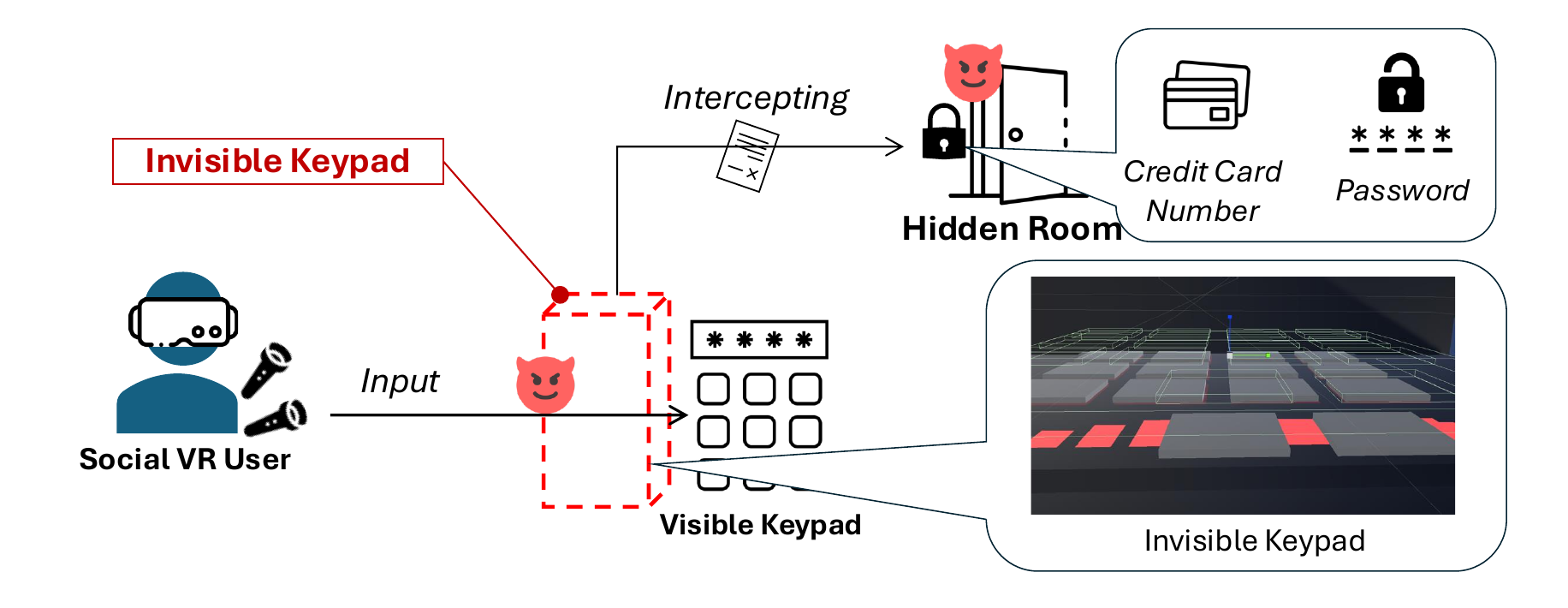}
    \caption{An in-world input hijack attack, where the adversary places an \emph{invisible object} on the virtual keypad to exfiltrate sensitive information to a \emph{hidden room} within a virtual world.}
    \label{fig:attack3}
\end{figure}

\begin{figure}
    \centering
    \subfloat[Before attack]{
        \includegraphics[width=0.4\linewidth]{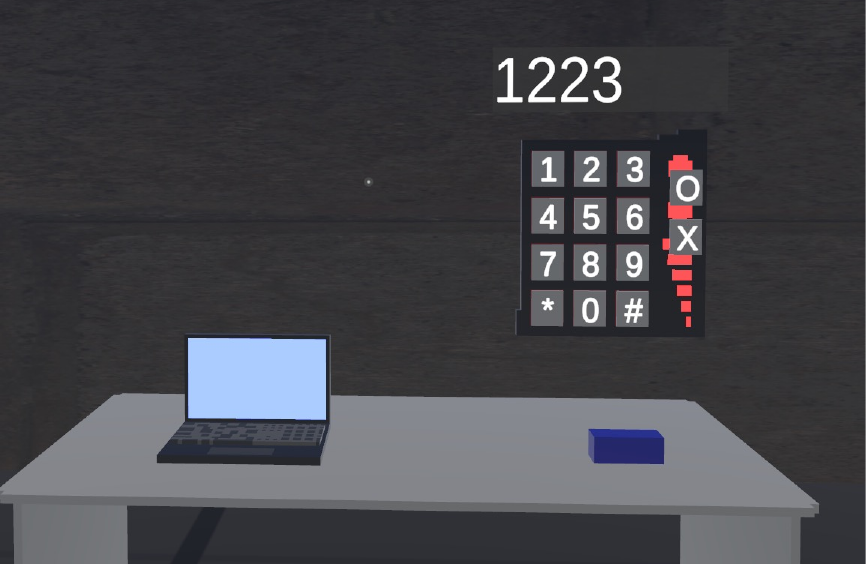}
        \label{fig:mitm_impl2}
    }
    \subfloat[After attack]{
        \includegraphics[width=0.4\linewidth]{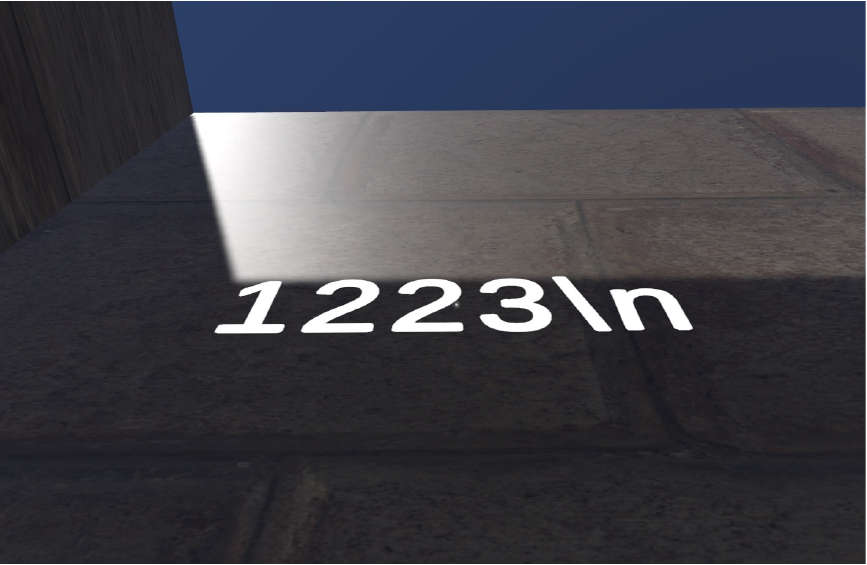}
        \label{fig:mitm_impl3}
        
    }
    \caption{In-world input hijack attack implementation.}
    \label{fig:mitm_impl}
\end{figure}

In this scenario, an adversary leverages an invisible object to exfiltrate users' sensitive information. Imagine a virtual shopping mall where users browse items and make purchases using a credit card.
The in-world payment system prompts users to enter credentials such as credit card numbers or passwords.
This attack targets the user's interaction with the payment interface, capturing input data through a hidden overlay positioned over the original interface. 
A prior study~\cite{cheng2024user} has demonstrated the feasibility of similar techniques in augmented reality (AR) environments. However, this attack becomes even more powerful in social VR platforms, where adversaries can exercise complete control over the virtual environment. Unlike AR, which is inherently constrained by the physical world, social VR enables adversaries to precisely place invisible objects and manipulate the environment to maximize user deception while minimizing the chances of detection.

Figure~\ref{fig:attack3} illustrates a scenario involving the in-world input hijack attack, and Figure~\ref{fig:mitm_impl} shows the corresponding scenes captured in our VRChat implementation. 
In this attack, an adversary constructs a shopping mall environment and places a virtual keypad to capture users' payment information. 
When a user enters sensitive data, an invisible object overlaid on the keypad intercepts their keystrokes. 
The adversary can employ a straightforward strategy to exfiltrate the captured information by transmitting user inputs to an external server via a covert channel. 
While this method enables immediate data collection, it increases the risk of detection by platform security mechanisms such as intrusion detection systems (IDS) or firewalls. 
Alternatively, we propose a new attack that redirects the intercepted input to a \emph{hidden room} within the virtual world. 
This method is more stealthy, as it avoids external network communication and keeps all exfiltration activities confined to the VR environment. 
This strategy is particularly effective in social VR platforms, where the adversary, acting as the world creator, can later access the hidden room and retrieve the exfiltrated data—potentially without alerting other users or platform moderators.

\vspace{0.05in}
\noindent\textbf{Attack Novelty.}
While prior work has demonstrated input interception through overlays in AR \cite{cheng2024user}, the novelty of our attack lies in its \emph{in-world exfiltration channel} rather than the fake interface itself. Instead of immediately transmitting captured input to an external server, the adversary redirects and stores the input within a hidden area of the virtual world that can later be accessed by the world creator. This attack design leverages a capability distinctive to social VR: an adversarial creator can control not only the interactive interface presented to users but also concealed objects within the same persistent virtual environment. The invisible overlay further improves stealth by preserving the appearance and expected behavior of the legitimate interface while covertly intercepting user input. Because the captured data remains within the virtual world for a while and requires no external network communication, conventional network-based monitoring cannot detect the exfiltration. Although we demonstrate the attack using an in-world payment interface, the same technique applies to other sensitive inputs handled by creator-controlled virtual interfaces.

\subsection{Avatar Quishing}

\begin{figure}
    \centering
    \includegraphics[width=\linewidth]{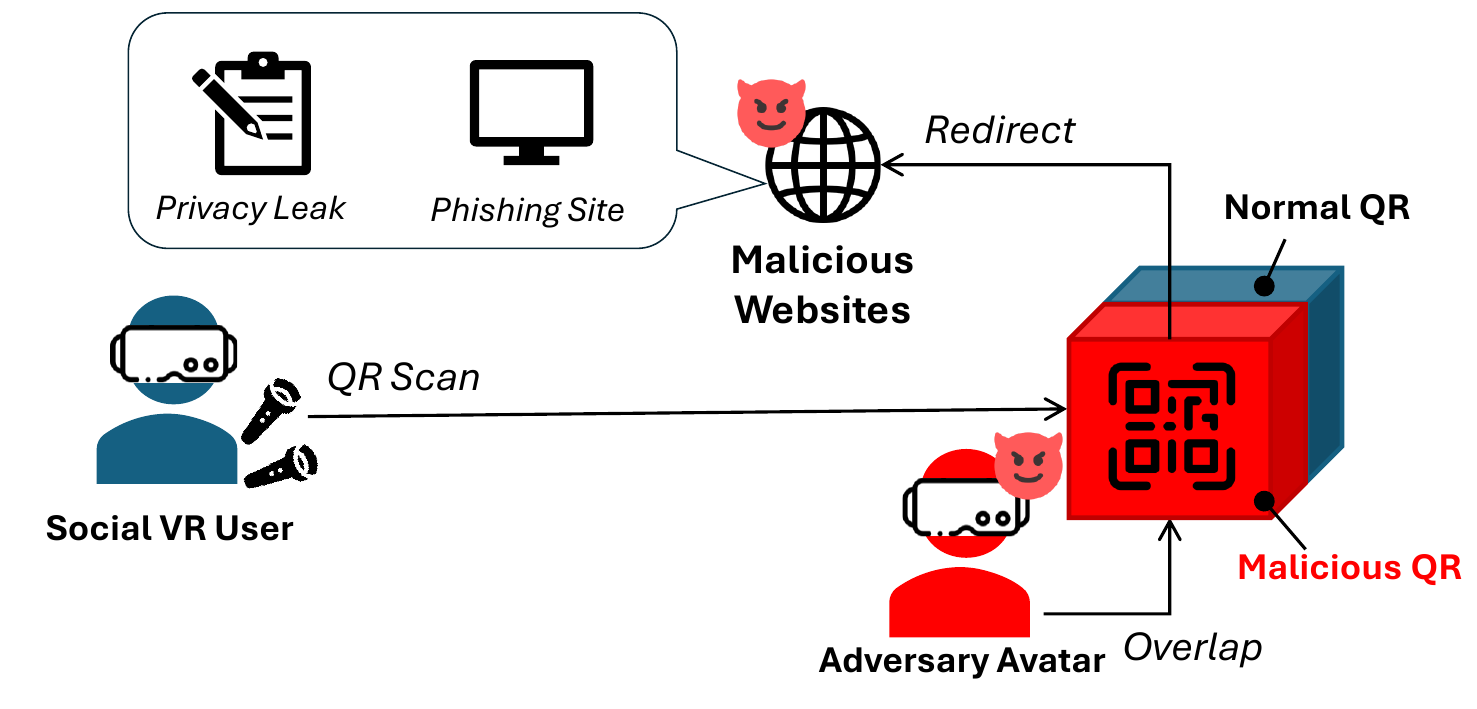}
    \caption{An avatar quishing attack where a \emph{malicious QR code} redirects user input to a malicious website.}
    \label{fig:quishing_attack1}
\end{figure}
\begin{figure}
    \centering
    \subfloat[An example of a VRChat avatar with a QR code that appears detached from the avatar's body.]{
        \includegraphics[width=.4\linewidth]{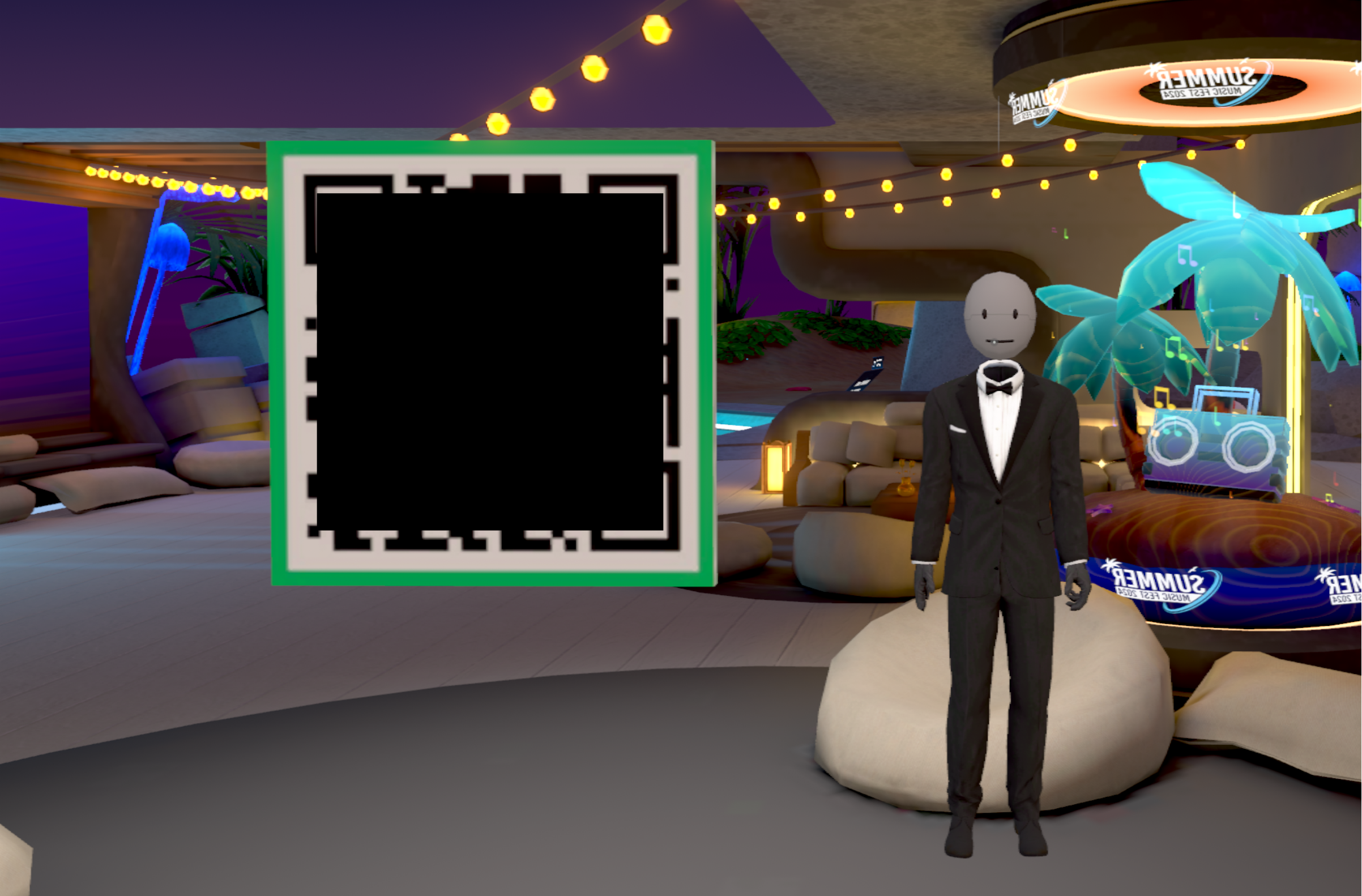}
        \label{fig:quishing_impl1}
        
    }
    \hspace{0.1in}
    \subfloat[Attack setting: the QR code A is benign while the QR code B is a malicious one attached to the adversary's avatar.]{
        \centering
        \includegraphics[width=.4\linewidth]{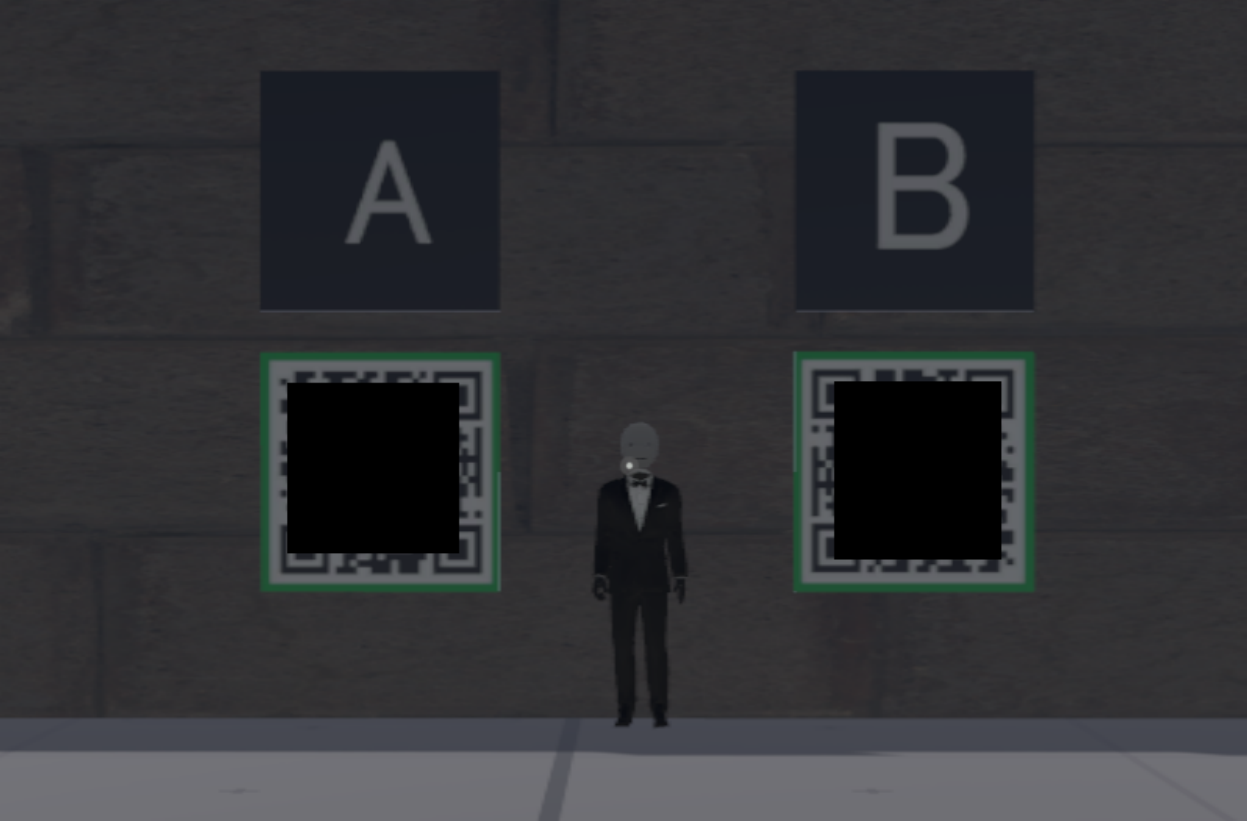}
        \label{fig:quishing_impl2}
        
    }
    \caption{Avatar quishing implementation.}
    \label{fig:quishing_impl}
\end{figure}

In this attack, an adversary embeds a malicious QR code into their customized avatar and positions the avatar so that the malicious code spatially overlaps an existing benign QR code in the virtual world. From the victim’s viewpoint, the malicious QR code visually replaces the benign one even though the underlying world content remains unchanged. This allows the adversary to introduce deceptive QR content into an independently created world without modifying the world itself.

In social VR platforms, avatars are often traded at high prices, with sellers showcasing their avatars in virtual worlds and attaching QR codes that link to purchase websites. QR codes are also commonly placed on avatars and in-world objects to direct users to external content, such as Discord servers or online storefronts, as reflected in a long-standing feature request for a native QR reader on VRChat's official feedback board~\cite{vrchat-feedback-qr}. However, prior studies on real-world QR code usage have shown that users rarely inspect the embedded URLs before visiting them~\cite{vidas2013qrishing,sharevski2024exploring}, making these QR codes an attractive vector for exploitation. Our examination of QR code scanner interfaces suggests that this risk also applies to social VR, as commonly used scanners provide limited opportunity to inspect URLs (see Appendix~\ref{appendix:qr} for more details). For example, VRQR~\cite{frostion-vrqr}---a QR code scanner widely used in VRChat---shows that its default setting opens the decoded URL automatically without requiring prior inspection by the user. Additionally, smartphones' QR code scanners, a common use case in tethered VR or Air Link setups, show only part of the URL, depending on configuration. This partial display can also be bypassed if an adversary uses a shortened or obfuscated URL. Accordingly, an adversary can mount a “Quishing” attack, wherein a malicious QR code redirects users to fraudulent or harmful websites. These sites may mimic legitimate services, prompting users to disclose sensitive information or unknowingly download malicious content---ultimately compromising their security.

While adversaries can embed malicious QR codes into various objects within a virtual world, we find that embedding the QR code directly into an avatar represents a particularly attractive variant. Platforms like VRChat support avatar creation and customization through a Software Development Kit (SDK)~\cite{vrchat_avatar}, which enables users to modify features such as skeletons, textures, and accessories. Leveraging the flexibility of this SDK, an adversary can subtly embed a malicious QR code into their avatar in a way that makes it appear visually detached from the avatar’s body---thereby avoiding easy detection by other users. Using this modified avatar, the adversary can enter any public virtual space (e.g., a shopping mall) and strategically position themselves in front of an existing benign QR code. This deliberate positioning causes the malicious QR code to fully overlap the legitimate one (as illustrated in Figure~\ref{fig:quishing_attack1} and implemented in VRChat in Figure~\ref{fig:quishing_impl}), causing other users' QR scanners to capture the attacker-controlled QR code and potentially directing the user toward a malicious destination.

Because QR scanning in social VR is often supported by user- or community-developed tools rather than platform-provided or security-validated scanners, safe handling of potentially malicious QR codes should not be assumed.\footnote{Desktop companion applications such as VRQR decode QR codes from VRChat screenshots. In our user study, participants used their own smartphones to scan the QR code displayed on the PC screen after viewing it in VR.} Accordingly, the risk should not be viewed solely as a scanner-side problem. Platforms can also mitigate such attacks at the content level by inspecting avatar assets for embedded QR codes and their suspicious destinations. We further discuss commonly used QR scanning applications and their limitations in Appendix~\ref{appendix:qr}.
\vspace{0.05in}

\noindent\textbf{Attack Novelty.} 
Unlike quishing attacks in the physical world, our approach uniquely leverages avatars---an essential and trusted component of social VR environments. 
By embedding a malicious QR code within an avatar, adversaries can effortlessly carry and position the attack vector, exploiting both spatial interactions and the inherent trust users place in avatars. 
This makes the deception significantly harder to notice compared to conventional object-based QR attacks.
\begin{figure}
    \centering
    \includegraphics[trim={0 5cm 0 12cm}, clip,width=0.8\linewidth]{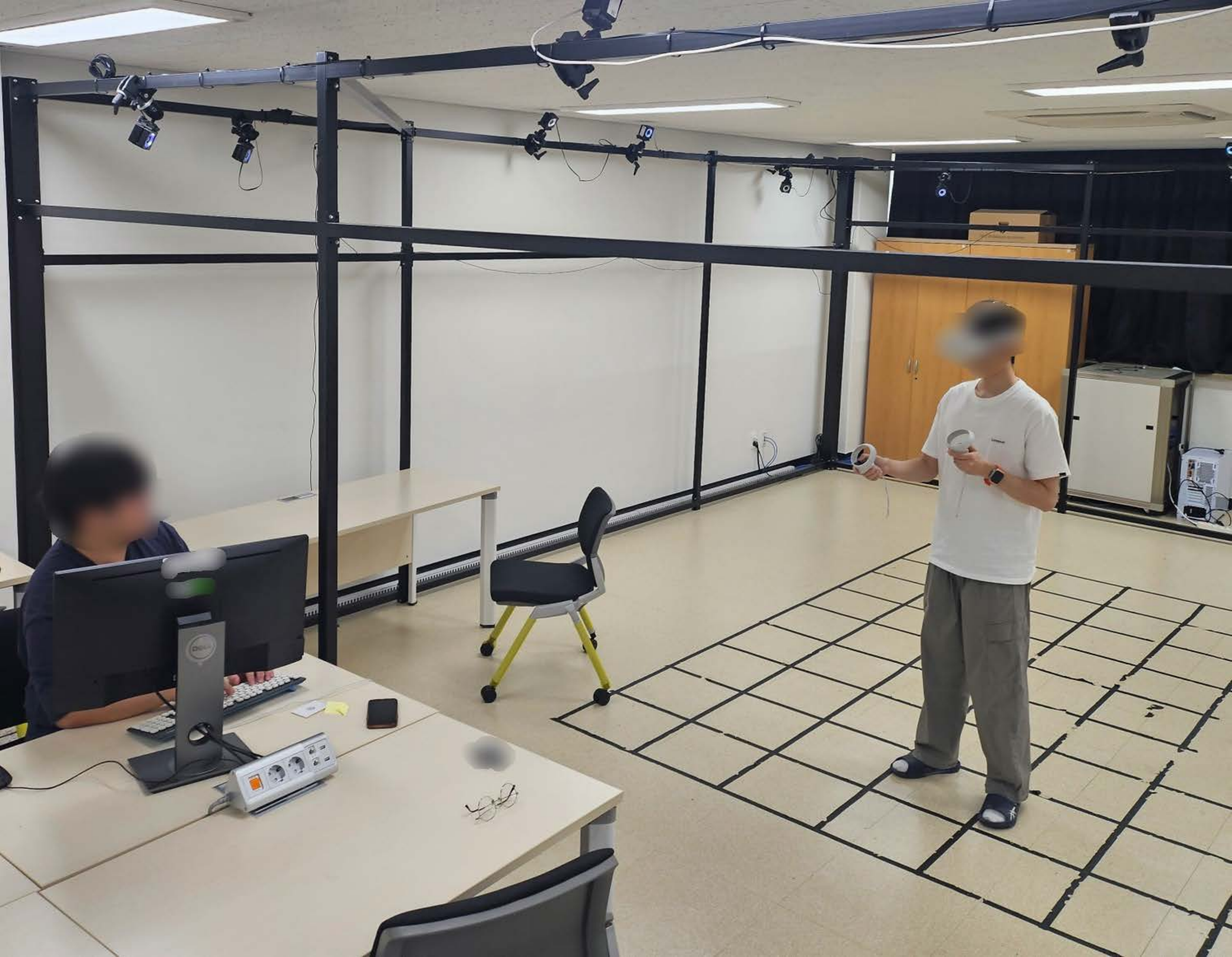}
    \caption{Experimental setup showing a participant (right) wearing a Meta Quest 2 headset and controllers, receiving instructions from the experimenter (left).}
    \label{fig:enter-label}
\end{figure}

\begin{figure}
    \centering
    \subfloat[Participants' prior VR usage]{
        \centering
        \includegraphics[width=0.47\linewidth]{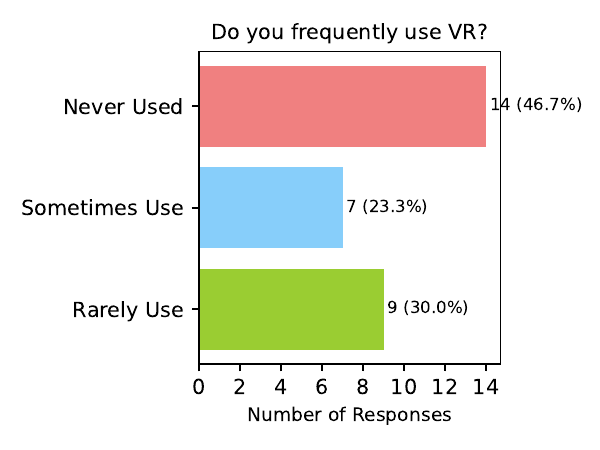}
        \label{fig:0-1}
    }
    \hfill
    \subfloat[Participants' gender distribution]{
        \centering
        \includegraphics[width=0.47\linewidth]{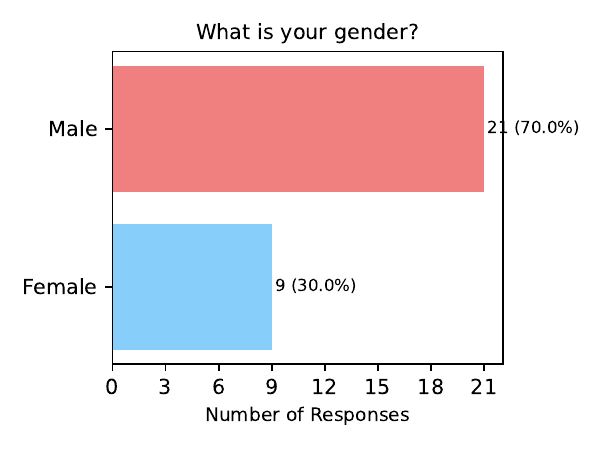}
        \label{fig:0-2}    }
    
    \subfloat[Participants' age distribution]{
        \centering
        \includegraphics[width=0.47\linewidth]{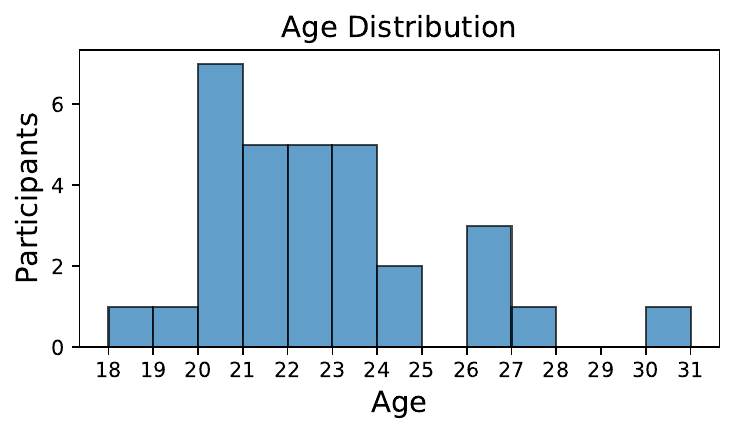}
        \label{fig:0-3}
    }
    \caption{Demographic information of participants and their familiarity with VR.}
    \label{fig:0}
\end{figure}

\section{User Study}
\label{section:user_study}

This section presents our IRB-approved user study involving 30 participants, demonstrating that our proposed attacks are effective within a proof-of-concept virtual world containing all four UI attacks. Only participants who could safely use a VR headset (e.g., no neurological conditions or known VR discomfort) were included in the study. The survey employed both open-ended and closed-ended items, with the type of each item specified in the respective figures. For the open-ended items, responses were categorized according to their core emotional expressions. Two researchers independently performed the coding, and inter-rater reliability was found to be perfect (Cohen’s κ = 1.0), indicating complete agreement between coders. No instances were observed in which respondents simultaneously expressed contradictory emotions.

\subsection{Experiment Setting}

We chose VRChat~\cite{vrchat} as the target social VR platform due to its popularity and well-established ecosystem for developing and publishing virtual content~\cite{vrchat_creator_guideline}. All attack scenarios were implemented using Unity 2022.3.6f1 in conjunction with the VRChat Creator Companion~\cite{vrchat_vcc}, the official SDK provided by VRChat. For the experiment, we used the Meta Quest 2 headset and recruited 30 participants from our institution, without imposing any specific demographic or technical restrictions. The Meta Quest 2 headset was connected wirelessly to a desktop PC through Air Link to run VRChat. To ensure physical safety, the experiment was conducted in a spacious, obstacle-free environment, securing a clearance of at least 6 meters in all directions (front, back, left, and right). Throughout the sessions, a researcher closely monitored the participants from a position allowing for immediate intervention to stop the experiment or offer advice if any potential risks were observed (see Figure~\ref{fig:enter-label}).

Regarding their familiarity with VR, nearly half of the participants (46.7\%, n = 14) reported that they had never used a VR headset before. The remaining 16 participants had prior VR experience, with 9 (30.0\%) reporting rare usage and 7 (23.3\%) reporting occasional usage. In terms of gender, the sample was predominantly male, with 21 male participants (70.0\%) and 9 female participants (30.0\%). Participants’ ages ranged from 18 to 31 years, with the majority between 20 and 24 years old.
Figure~\ref{fig:0} presents a breakdown of participants’ familiarity with VR headsets, as well as their gender and age distributions. 
To assess the extent to which users were aware of the attacks, participants were not informed of the malicious nature of the virtual environment prior to the study.

To evaluate the proposed attacks, we developed a malicious virtual world comprising four distinct rooms, each designed to reflect common social VR activities such as exploring virtual shops, engaging with mini-games, or claiming localized content. Participants progressed through these rooms in a predefined sequence and were instructed to interact with the available objects to complete the task given in each room. These tasks were embedded within immersive environments that closely resemble typical user interactions on social VR platforms, allowing us to observe participant responses to the deceptive elements under realistic conditions. Participants were only informed that they would be taking part in a normal social VR experience. Before starting the tasks, they were given a 10-minute acclimation period to freely explore the environment and familiarize themselves with object interactions and VR controller usage. This phase ensured that participants became comfortable navigating the virtual space and operating the VR device.

Our study adopts a within-subjects design in which each participant completed all four attack scenarios (E1--E4) in a fixed order. The study was presented as a guided walkthrough introducing the basic interactions of the virtual world, so the scenario order was fixed to provide a natural tutorial-style experience. To reduce potential order and learning effects, each room featured a perceptually distinct task, including stacking tomatoes, a memory game, a purchase, and a QR scan. Survey responses showed no evidence of learning effects across the four scenarios.

After completing the tasks, participants were fully debriefed and informed about the true purpose of the study, including a detailed explanation of how their actions were influenced or exploited in each scenario. Following their participation, each individual completed a questionnaire (see Appendix~\ref{appendix:questionnaire}) during which the actual purpose of the experiment was disclosed. To ensure participants could accurately recall the details of each experiment, the questionnaire included reference screenshots of the corresponding rooms and scenarios. We applied binomial tests to binary questions and chi-square goodness-of-fit tests to multi-category questions to evaluate deviation from uniform responses.

\vspace{0.05in}

\noindent\textbf{Ethical Consideration.} We identify two potential ethical risks: (i) the risk of involving human subjects in our user study and (ii) the risk of offending innocent VR users with our attack experimentation. For this, we carefully devised the protocol to prevent any discomfort or injury during our user experiment. For example, every participant is assisted to have a good fit with the VR headset in order not to experience VR sickness or discomfort. Note that our study has been granted IRB approval from our institutional review board. Last but not least, our attack-crafted virtual world has remained \emph{private}, meaning that only invited users (i.e., our experiment participants) can join; unsolicited access to a private world is prohibited by VRChat. Note that our experiments did not collect any private information from participants during either the experiment or the questionnaire. We reported the deceptive UI attacks identified in this study to VRChat through its security reporting channel on November 10, 2026. Below, we describe implementation details for the experiments (\textbf{E1}-\textbf{E4}) and their results along with users' responses.

\subsection{E1. Layered Click Redirection}
\label{subsec:e1_clickjacking}

\noindent\textbf{Attack Implementation.}
The room contains two green buttons arranged front and back with desks, as shown in Figure~\ref{fig:click_impl1}. A participant clicks the first green button, which makes tomatoes appear on the desk. Next, the participant clicks the second button, expecting the same result (i.e., tomatoes popping up). However, we strategically place two buttons—the clickjacking object and the decoy object—in the same position, exploiting the overlapped object strategy (Section~\ref{subsec:clickjacking_design}). Since the decoy still generates tomatoes, participants are unlikely to notice anything unusual\footnote{The specific version of Unity used is critical, as the rendering order of overlapping objects may vary across versions. In some versions, this overlap results in flickering artifacts, reducing the reliability of the attack. However, the rendering order can be controlled by adjusting the RGB values of the Unity Material component; objects with higher blue and lower red values tend to be rendered on top.}.  Leveraging this behavior, we set the clickjacking object to red and the decoy object to green, ensuring the clickjacking object remains concealed.

Within the clickjacking object, we embedded a YouTube URL because the VRChat programming language, Udon, has a video player allowlist for URLs~\cite{vrchat_allow_list}; thus, only specific trusted domains are allowed. However, if users enable the ``Allow untrusted URLs'' option, URLs from any network domain are allowed. Note that this option is occasionally turned on for third-party applications or in-VR advertisements~\cite{lee2021adcube} to work properly.
Consequently, an adversary embedding a custom URL can trick users into playing a malicious video or artificially boosting views on their YouTube channel---a form of ad fraud~\cite{lee2021adcube}.

\begin{figure}
    \centering
    \subfloat
    [Perceived difference between two buttons ($p<.001$, $g=0.33$, Closed-ended)]
    {
        \centering
        \includegraphics[width=.47\linewidth]{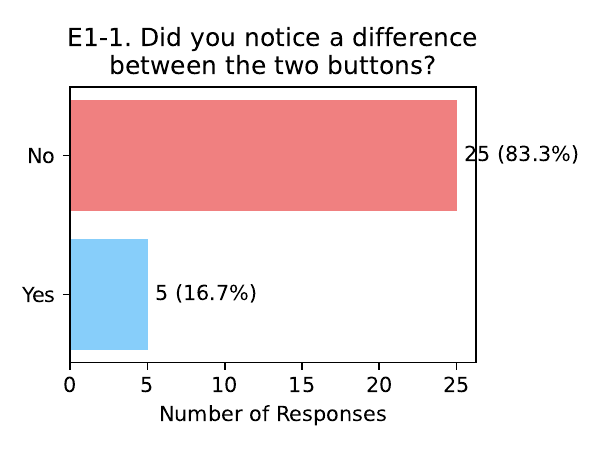}
        \label{fig:1-1}
    }
    \hfill
    \subfloat
    [Reasons for reporting no difference ($\chi^2(3)=30.20$, $p<.001$, $w=1.10$, Open-ended)]
    {
        \centering
        \includegraphics[width=.47\linewidth]{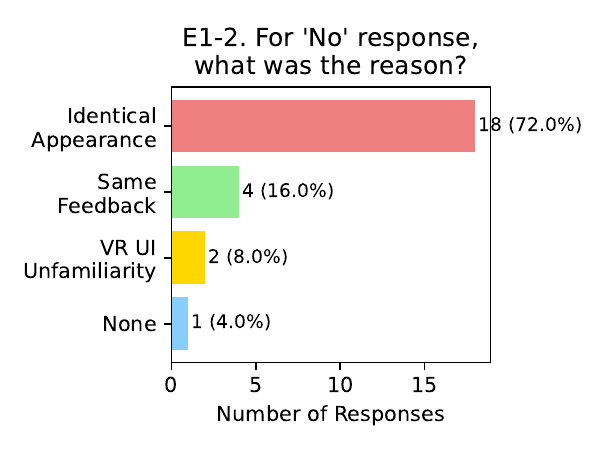}
        \label{fig:1-2}
    }
    \vspace{-0.1in}
    \subfloat
    [Expected behavior on second button ($\chi^2(2)=34.20$, $p<.001$, $w=1.07$, Open-ended)]
    {
        \centering
        \includegraphics[width=.47\linewidth]{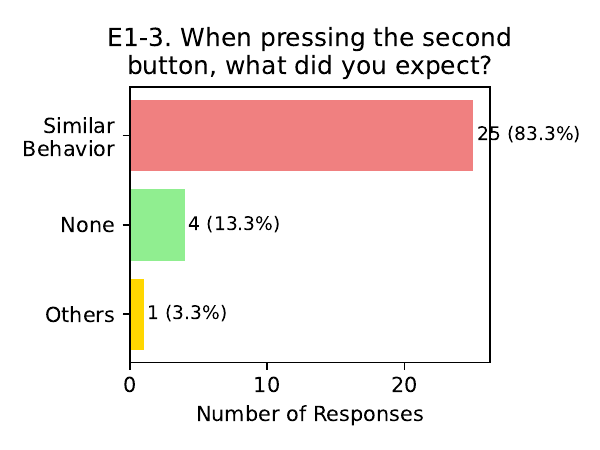}
        \label{fig:1-3}
    }
    \hfill
    \subfloat
    [Emotions when video played ($\chi^2(4)=12.00$, $p=.017$, $w=0.63$, Open-ended)]
    {
        \centering
        \includegraphics[width=.47\linewidth]{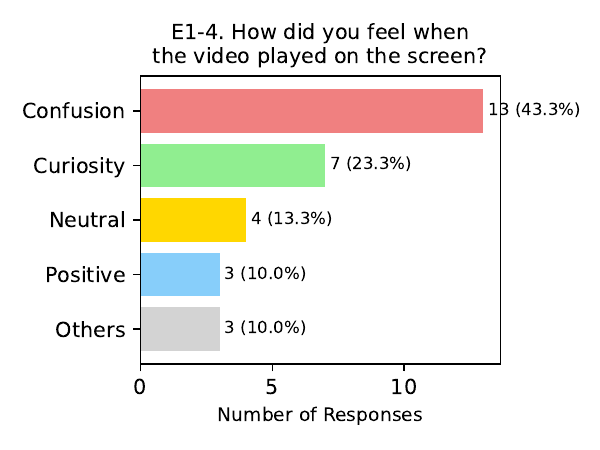}
        \label{fig:1-4}
    }
    \caption{Summary of user responses from the layered click redirection experiment (E1).}
    \label{fig:1}
\end{figure}

\vspace{0.05in}

\noindent\textbf{Task Design.} In this experiment, participants were instructed to stack 20 tomatoes on each of the two tables. Participants could spawn tomatoes by clicking a button and then manually place them on the table. This task was introduced as a simple adaptation exercise. However, in reality,
clicking the second button triggers the clickjacking object, which plays a video (see Figure~\ref{fig:click_impl2}), leading to an unexpected result for the user.

\vspace{0.05in}
\noindent\textbf{Results.} Figure~\ref{fig:1} summarizes users' responses in Experiment 1. 
We observed that most participants (83.3\%) did not notice any difference between the two buttons before pressing them, as confirmed by a binomial test ($p < .001$, Cohen’s $g = 0.33$). 
The main reasons cited were the identical appearance of the two buttons (72\%) or the similar feedback they provided (16\%), as supported by a chi-square goodness-of-fit test ($\chi^2(3) = 30.20$, $p < .001$, Cohen’s $w = 1.10$).
Additionally, a majority of participants (83.3\%) expected the second button to behave similarly to the first one (i.e., to produce tomatoes), as supported by a chi-square goodness-of-fit test ($\chi^2(2) = 34.20$, $p < .001$, $w = 1.07$).
When the video played, participants reported feeling confusion (43.3\%) and curiosity (23.3\%), as supported by a chi-square goodness-of-fit test ($\chi^2(4) = 12.00$, $p = .017$, $w = 0.63$).
For example, one participant stated:
\begin{quote}
    \textit{``I was confused why it showed up, and thought it was not a result of my action.''}
\end{quote}
Another participant expressed:
\begin{quote}
    \textit{``I was confused because the expected item didn't show up. Since nothing changed after pressing the button, I thought I might have done something wrong.''}
\end{quote}
When dividing the 30 participants into those with prior VR experience and those without, we found that 3 participants with VR experience and 2 participants without VR experience noticed the attack. A chi-square test revealed no significant difference between the groups ($p \approx .74 > .05$), suggesting that our attack was perceived as equally strong regardless of prior VR experience. These results suggest that social VR users strongly rely on visual appearance when predicting UI actions of a target object and our attack is effective by exploiting this behavior.

\subsection{E2. Raycasting Phishing}
\label{subsec:e2_raycasting}

\noindent
\textbf{Attack Implementation.} In this room, users initially interact with blue buttons on the wall via raycasting, as illustrated in Figure~\ref{fig:dos_imp1}. However, a red button (Figure~\ref{fig:dos_impl2}) serves as a trigger object that tracks the user’s position and deploys invisible walls around them. Consequently, users can no longer click the blue buttons; as shown in Figure~\ref{fig:dos_impl3}, the absence of a mouse icon over the button indicates it is no longer clickable. Despite this, users remain unaware of the attack. Meanwhile, a banner appears displaying fake information alongside a QR code\footnote{The QR code is masked to maintain anonymity.} (Figure~\ref{fig:dos_impl4}). This banner misleads users into believing they are experiencing a software bug or hardware malfunction, prompting them to scan the adversary’s QR code for a supposed solution.

\vspace{0.05in}

\noindent\textbf{Task Design.} The task in this experiment was a memory task involving seven food items. Participants were instructed to press seven blue buttons in sequence to view different foods and memorize them. Once ready, they would press a red button to begin the test and attempt to reproduce the sequence by pressing the same blue buttons again. However, pressing the red button triggered an invisible object to be attached to each participant, preventing them from interacting with the blue buttons.

\begin{figure}
    \centering
    \subfloat[Noticed changes after red button ($p<.001$, $g=0.37$, Closed-ended)]{
        \centering
        \includegraphics[width=0.47\linewidth]{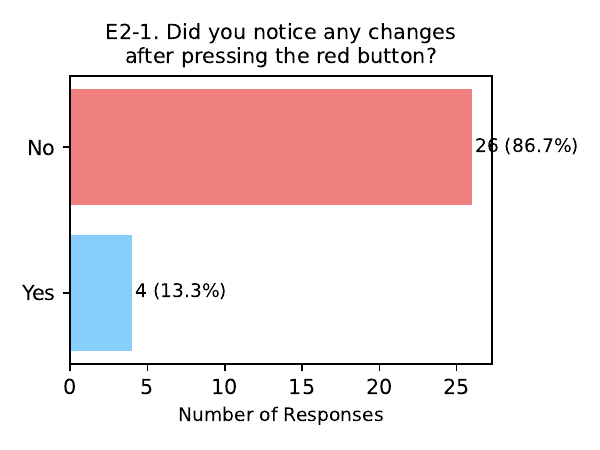}
        \label{fig:2-1}
    }
    \hfill
    \subfloat[Cause of button stopping ($\chi^2(4)=21.67$, $p<.001$, $w=0.85$, Closed-ended)]{
        \centering
        \includegraphics[width=0.47\linewidth]{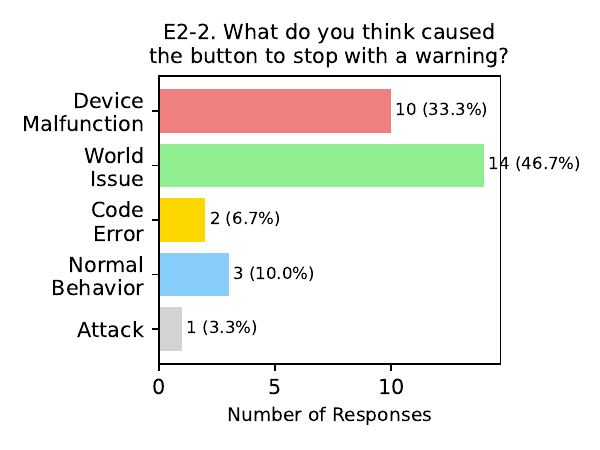}
        \label{fig:2-2-bar}
    }
    \vspace{-0.1in}
    \subfloat[Considered scanning QR code ($\chi^2(2)=13.52$, $p=.001$, $w=0.68$, Closed-ended)]{
        \centering
        \includegraphics[width=0.47\linewidth]{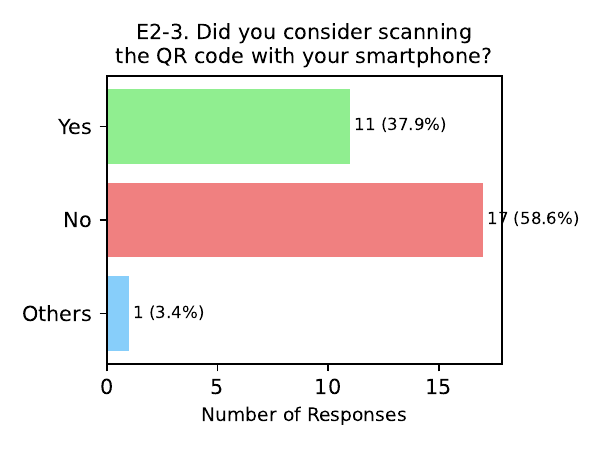}
        \label{fig:2-3}
    }
    \vspace{-0.1in}
    \caption{Summary of user responses from the raycasting phishing experiment (E2).}
    \label{fig:2}
\end{figure}

\vspace{0.05in}
\noindent\textbf{Results.} Figure~\ref{fig:2} presents the participants' responses for Experiment 2. 
Most participants (86.7\%) did not notice any changes—such as the presence of an invisible wall—after triggering the red button, except for the fake UI, as confirmed by a binomial test ($p < .001$, $g = 0.37$).
Interestingly, many participants attributed the issue to a problem with the virtual world (46.7\%) or a device malfunction (33.3\%) when they discovered that the blue button was unclickable, as supported by a chi-square goodness-of-fit test ($\chi^2(4) = 21.67$, $p < .001$, $w = 0.85$).
For instance, one participant commented:
\begin{quote}
    ``\emph{As I'm not familiar with VR, I thought there might be a device problem.}''
\end{quote}
More concerning is that some participants (37.9\%) expressed a willingness to scan the QR code shown on the fake banner, as supported by a chi-square goodness-of-fit test ($\chi^2(2) = 13.52$, $p = .001$, $w = 0.68$). When dividing the 30 participants into those with prior VR experience and those without, we found that 1 participant with VR experience and 3 participants without VR experience noticed the attack. A chi-square test revealed no significant difference between the groups ($p \approx .22 > .05$), suggesting that our attack was perceived as equally strong regardless of prior VR experience.
These results highlight how adversaries can easily mislead users by manipulating raycasting—a new and unfamiliar UI feature in VR for many users.

\subsection{E3. In-World Input Hijack}
\label{subsec:e3_in-world}

\noindent\textbf{Attack Implementation.} In this experiment, we developed a mini VR mall where users can explore and purchase items. We simulated a payment scenario by creating a virtual keypad for password input (Figure~\ref{fig:mitm_impl2}). A transparent duplicate keypad was constantly positioned in front of the original, intercepting user input. The duplicate keypad contains an embedded script that records input and stores it on the display located inside the hidden room (Figure~\ref{fig:mitm_impl3}). The intercepted data is stored in the form of text-based pairs of user credentials and passwords. An adversarial world creator could access this room at any time to view the intercepted data.

\vspace{0.05in}

\noindent\textbf{Task Design.} The task involved selecting and purchasing a preferred item from a set of six products. Participants were asked to set a personal password in advance, which would be required to complete the purchase. During this task, the system covertly harvested the participant’s password, and at the end of the experiment, participants were informed about the exposure of their personal data recorded in the hidden room.

\begin{figure}
    \centering

    \subfloat[Emotional reaction to keypad data redirection ($\chi^2(4)=9.33$, $p=.053$, $w=0.56$, Open-ended)]{
        \centering
        \includegraphics[width=0.46\linewidth]{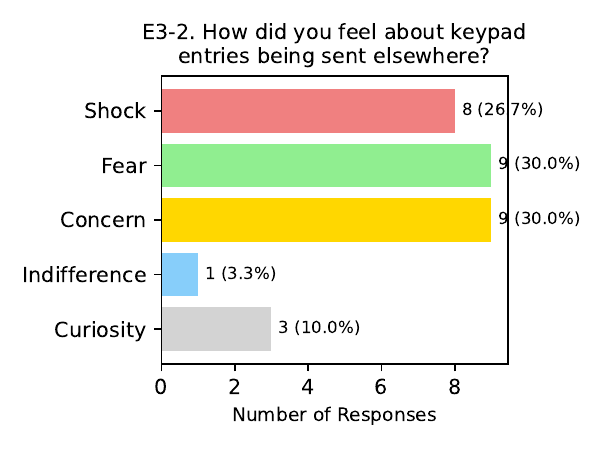}
        \label{fig:3-2}
    }
    \hspace{0.05in}    
    \subfloat[Guessed implementation of redirection ($p < .001$, $g=0.33$, Open-ended)]{
        \centering
        \includegraphics[width=0.46\linewidth]{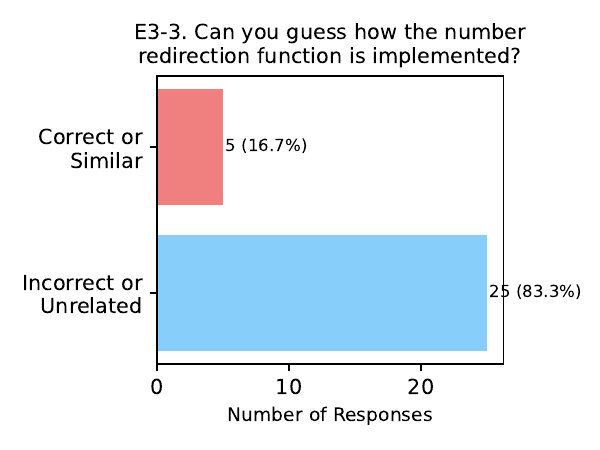}
        \label{fig:3-3}
    }
    \caption{Summary of user responses from the in-world input hijack attack experiment (E3). Note that no participant noticed anything unusual about the keypad (E3-1).}
    \label{fig:3}
\end{figure}

\vspace{0.05in}

\noindent\textbf{Results.} Figure~\ref{fig:3} presents the users' responses in Experiment 3. As the transparent duplicate keypad preserved the appearance and behavior of the original keypad, all participants (100\%) perceived the interface as fully legitimate while typing on the keypad. After revealing that their inputs had been redirected to a secret room, we asked them about their reactions. Participants expressed concern (30\%), fear (30\%), and shock (26.7\%), although this distribution did not reach conventional statistical significance ($\chi^2(4) = 9.33$, $p = .053$, $w = 0.56$). For example, one participant remarked:
\begin{quote}
    ``\emph{It was unsettling to realize how easily a cracking could happen in a virtual space. I think most people, who are not familiar with VR, will become easy attack targets.}''
\end{quote}
Another participant commented:
\begin{quote}
    ``\emph{I would definitely be victimized without realizing it if this were a real attack.}''
\end{quote}
We also asked participants if they could guess how the redirection function was implemented. The majority (83.3\%) provided incorrect or unrelated responses, as confirmed by a binomial test ($p < .001$, $g = 0.33$). This group included participants who had experience developing VR content through coursework. Together with the finding that all participants perceived the keypad as legitimate, these responses suggest that the transparent duplicate keypad was difficult to recognize as a possible mechanism for intercepting input. These findings suggest that such redirection attacks in social VR environments could leave users vulnerable to information leakage without their awareness.

\begin{figure}
    \centering
    
    \subfloat[Doubted authenticity of QR code ($\chi^2(2)=34.20$, $p<.001$, $w=1.07$, Closed-ended)]{
        \centering
        \includegraphics[width=0.48\linewidth]{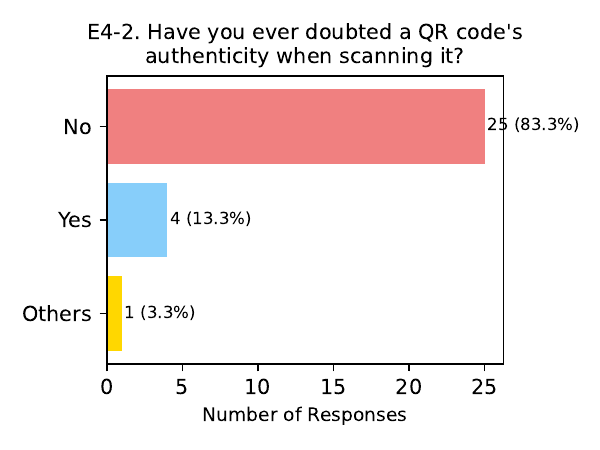}
        \label{fig:4-3}
    }
    \caption{Summary of user responses from the avatar quishing experiment (E4). Note that all participants (100\%) reported no visual difference between the two QR codes (E4-1).}

    \label{fig:4}
\end{figure}

\subsection{E4. Avatar Quishing}
\label{subsec:e4_quishing}

\noindent
\textbf{Attack Implementation.} For this experiment, we downloaded a publicly available free avatar and attached a malicious QR code next to it. We then successfully uploaded this avatar to the actual VRChat platform, confirming that such malicious avatars containing QR codes can be uploaded without any platform-level restrictions. Although the uploaded malicious QR code did not contain harmful content, it was associated with a URL flagged as an unsafe domain by Google’s Safe Browsing API. The QR code appears detached from the avatar’s body (Figure~\ref{fig:quishing_impl1}), making it difficult for users to distinguish it from other objects. We then placed two identical QR codes, A and B, in a room and positioned the adversary’s avatar so that its malicious QR code completely covered QR code B (Figure~\ref{fig:quishing_impl2}).

\vspace{0.05in}

\noindent
\textbf{Task Design.} Participants were asked to scan a QR code placed within the room. In our Air Link setup, participants removed the headset and scanned the QR code on the PC screen with their smartphones. After viewing the QR code in VR, participants used their own smartphones to scan the code displayed on the PC screen and access the questionnaire (refer to Appendix~\ref{appendix:qr} to see how the QR code appears in the various QR code scanners). The QR codes directed participants to two different versions of a debriefing survey that revealed the true research focus on deceptive UI attacks and explained how each task demonstrated specific security threats. Although both codes appeared to belong to the virtual world, one was attached to an avatar we deployed to act as a research assistant and covered QR code B. The questionnaire asked whether participants noticed any visual difference between the two QR codes, excluding differences in the QR pattern itself. This question assessed their perception of the deceptive QR placement rather than their ability to identify a malicious destination from the decoded URL.

\vspace{0.05in}

\noindent\textbf{Results.} Figure~\ref{fig:4} shows participants' responses from Experiment 4. 
All 30 participants (100\%) reported that they did not notice that an avatar-attached QR code was covering the original QR code.
Upon revealing that one QR code could lead to a malicious site, participants expressed surprise. 
One participant remarked: 
\begin{quote}
    ``\emph{I scan QR codes every day in my life. It is terrifying that they could be manipulated by overlapping.}''
\end{quote}
All participants (100\%) had prior experience using QR codes, likely due to their use for attendance checks or bicycle-sharing systems. 
A majority (83.3\%) admitted they never questioned the authenticity of a QR code before scanning,  as supported by a chi-square goodness-of-fit test ($\chi^2(2)=34.20$, $p<.001$, $w=1.07$).
One participant noted:
\begin{quote}
    ``\emph{I never realized that scanning QR codes or clicking links, which I routinely did without suspicion, could potentially lead to personal information theft.}''
\end{quote}
These findings suggest that QR-code overlays can be difficult to notice. As illustrated in Appendix~\ref{appendix:qr}, the information available to users before opening a QR-code destination varies across scanning applications and settings. Prior work has demonstrated users' difficulties with deceptive URLs~\cite{lain2025url} and phishing webpages that closely resemble legitimate websites~\cite{ji2025evaluating}. Users may therefore fail to recognize an unintended destination without carefully inspecting both the decoded URL and the webpage. Combined with Avatar Quishing, these deceptive techniques may make the redirection even harder to detect.

\section{MetaScanner}
\label{sec:metascanner}

In this section, we introduce \ourtool{}, a practical countermeasure against the proposed UI attacks. \ourtool{} aims to proactively mitigate these threats by scanning and detecting malicious objects or scripts within virtual content. This allows social VR platform providers and users to verify that uploaded or downloaded virtual content is free of malicious logic.

\subsection{Design Considerations}

The development of \ourtool{} is guided by several key design considerations (DC):

\vspace{0.05in}
\noindent\textbf{DC1. Proactive and Lightweight Scanner.} Virtual content on social VR platforms often contains numerous objects and scripts, resulting in a substantial volume of virtual content. For example, VRChat hosts approximately 25,000 published virtual worlds~\cite{vrchat}. A naive approach is to dynamically analyze the objects and scripts within all this content by running them, which is time-consuming and introduces significant overhead. Therefore, \ourtool{} must be lightweight to enable social VR platform providers to proactively analyze a large number of uploaded virtual worlds and repeatedly iterate the analysis as content gets updated.

\vspace{0.05in}

\noindent\textbf{DC2. Supporting Platform-Specific Policies.} The criteria for identifying suspicious objects and scripts differ across social VR platforms. For instance, VRChat and Rec Room prohibit directly opening URLs via script APIs, as per platform security policies. Conversely, Mozilla Hubs allows users to host worlds and servers from their own machines and share them through WebXR tools, without restrictions on world content. This openness permits the inclusion of scripts that open URLs within worlds. To address these variations, \ourtool{} must support policies that define which objects and scripts are allowed or disallowed according to each platform's specific rules.

\vspace{0.05in}

\noindent\textbf{DC3. Preventing Modified Library Linking.} In addition to detecting suspicious objects and scripts, \ourtool{} must also assess the integrity of libraries that a virtual world package depends on. This is crucial to counter the threat of adversaries covertly injecting malicious logic into third-party libraries, akin to supply chain attacks~\cite{ladisa2023sok}. To mitigate this risk, \ourtool{} verifies the integrity of these libraries by comparing them with their original versions. This process ensures that uploaded content does not depend on modified libraries.

\begin{figure}
    \centering
    \includegraphics[width=\linewidth]{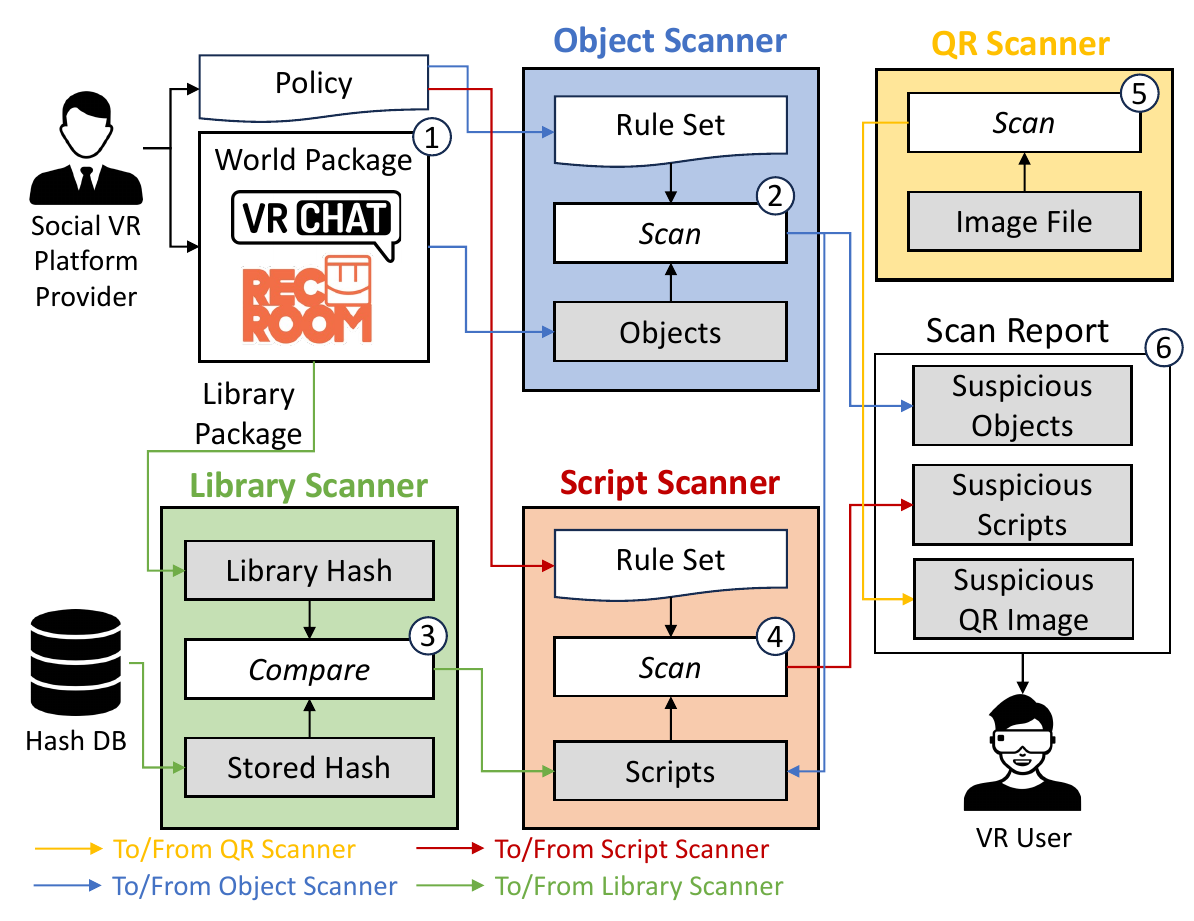}
    \caption{An architectural overview of \ourtool{}.}
    \label{fig:metascanner}
\end{figure}
\subsection{System Design}

\noindent\textbf{Overview.} Figure~\ref{fig:metascanner} shows the architectural overview of \ourtool{}, which is designed to statically analyze virtual worlds provided by social VR platforms to detect suspicious objects and scripts. \ourtool{} is composed of three submodules: (i) an object scanner, (ii) a library scanner, and (iii) a script scanner. The process begins with \ourtool{} receiving \emph{policies} from the social VR platform provider\footnote{We assume the social VR platform provider is a trusted entity whose goal is to maintain a safe environment for its users. Therefore, \ourtool{} is designed to be deployed and operated by the provider as a proactive defense mechanism. We exclude scenarios where the platform provider itself acts maliciously or where the core platform infrastructure is compromised.} that define the suspicious objects and scripts to suppress in order to prevent UI attacks. These policies are then translated into \emph{rule sets} for each module. \circled{1}~\ourtool{} first processes the package file of a target virtual world and applies the modules to analyze its objects and scripts. \circled{2}~The object scanner checks whether the package file contains suspicious objects specified in the rule set and logs any matches. \circled{3}~It also verifies the integrity of the package's libraries by comparing their hash values against pre-calculated values from trusted libraries to ensure they haven't been tampered with. \circled{4}~The script scanner then inspects scripts from both the target virtual world and any libraries that failed the integrity check. \circled{5}~Running independently, the QR scanner examines image files in the world package to detect QR codes and identify their associated URLs. \circled{6}~If any scripts use suspicious APIs as defined by the rule set, they are flagged. \circled{7}~Finally, \ourtool{} generates a report, summarizing the number of suspicious objects, scripts, and QR codes found. Based on this analysis, \ourtool{} provides a warning, allowing social VR users to make an informed decision about entering the virtual world. Next, we examine each module in detail.

\vspace{0.05in}
\noindent\textbf{Object Scanning.} 
The goal of the object scanner is to analyze suspicious objects within a virtual world. We define \emph{suspicious objects} based on our proposed attacks, which include transparent objects and overlapped objects, as previously described (Section~\ref{section:ui_attacks}). Our experiment results (Section~\ref{subsec:scan_effect}) indicate that worlds using transparent objects are extremely rare. Another case involves overlapped objects with the same scale and prefabs that are aligned within a certain radius. From our collected world data, we found no instances where objects with the same scale and prefabs were positioned within close proximity (overlapping or nearly overlapping). While it is common to use the same scale and prefabs, such objects are not typically placed in extremely close proximity, leading us to define these as suspicious objects intended for clickjacking or object-in-the-middle attacks. 

Finding such suspicious objects involves two main steps: First, \ourtool{} identifies \emph{Components}, which represent the properties or elements of a \emph{GameObject}\footnote{In Unity world packages, objects are referred to as \emph{GameObjects}, and their functional pieces are referred to as \emph{Components}.}, such as physical characteristics, location, and specific behaviors. For example, a content creator can create a transparent object by setting the \texttt{alpha} value to 0. Since these properties are often configured through scripts, it is crucial to thoroughly examine them at this stage.  Second, \ourtool{} checks the scripts attached to each Component. Since world developers can add scripts to Components, this step is critical because any of these scripts can be triggered when a user interacts with the object. The gathered information is then passed to the script scanner for a detailed analysis of the scripts' internal logic. Note that we rely on the Unity engine to identify these objects and components.

\vspace{0.05in}

\noindent\textbf{Library Scanning.}
In Unity, developers can utilize a wide range of libraries provided by Unity and third-party developers. These third-party libraries are publicly accessible and available for download from the Unity Asset Store~\cite{unity_asset_store}. However, these libraries are also susceptible to supply chain attacks~\cite{ladisa2023sok}. While our proposed attack scenarios do not explicitly focus on library poisoning, we consider it a realistic threat in social VR environments; therefore, ensuring the integrity of these libraries is crucial. Since scrutinizing every library's code manually is time-consuming and resource-intensive, we propose an alternative approach. The library scanner computes a hash value for each script within a library and stores it in a database. This process is repeated for all published libraries. Then, when the library scanner receives a world package, it calculates the hash value for the library within that package. If there is a mismatch between the stored and calculated hash values, it indicates that the scripts have been modified. In such cases, the scripts are forwarded to the script scanner for further analysis. Our evaluation results demonstrate that the Library Scanner effectively identifies library-level tampering, successfully detecting contaminated libraries regardless of whether their hash values were registered in advance.

\vspace{0.05in}

\noindent\textbf{Script Scanning.}
The script scanner analyzes suspicious scripts identified by the previous two processes using \emph{static analysis} to avoid the substantial overhead associated with executing these scripts in Unity, which would necessitate object creation and interaction handling. The scanner specifically uses string matching to detect potentially harmful APIs, such as \texttt{Application.OpenURL}~\cite{unity_url}, which opens a specified URL when triggered, and \texttt{System.Net.Sockets}, which establishes a socket for external communication. Scrutinizing these APIs is crucial, as they can be exploited to launch information leakage attacks (e.g., object-in-the-middle attacks).

\vspace{0.05in}

\noindent\textbf{QR Scanning.}
The QR scanner is responsible for finding QR codes that have suspicious URLs within virtual content. In Unity, inserting a QR code typically involves adding a mapping image to a Material. As QR codes are generally in image file format, the corresponding image files must reside within the Asset folder. Thus, the QR scanner begins by scanning image files within the Asset folder. Once an image file is detected, it is decoded using a QR decoder to determine whether the content resembles a URL. If the decoded content is in URL format, the image is flagged as a ``Suspicious QR Image,'' and a notification is sent to the user. %
\section{Evaluation}

\subsection{Implementation}
We implemented \ourtool{} using the Unity Engine, leveraging C\# as the primary programming language. Unity provides several built-in functions, such as \verb|FindObjectsOfType| and \verb|GetComponents|, which facilitate efficient object scanning and component inspection. These functions are particularly advantageous due to their speed and ease of integration into the scanning workflow. To simplify the policy enforcement, we utilized Unity's component structure, where each component's \verb|.name| property stores a string value. This allowed for straightforward string matching against predefined policies. For deeper inspections, such as retrieving type and value information, we employed reflection methods like \verb|GetProperty|, \verb|GetType|, and \verb|GetValue|. For QR code scanning, we integrated Zebra Crossing \cite{zxing}, an open-source library provided by Google, enabling robust QR code detection and decoding capabilities.

The library scanning was implemented to discover the varied locations of package installations. Libraries can exist within the \verb|Asset| folder or be managed through the Unity Package Manager, which stores packages in the user's \texttt{Cache.package} directory. In such cases, the Asset folder contains a manifest file detailing the packages used by the project. Our implementation prioritizes scanning libraries found directly within the Asset folder since these can be modified by the user. If a library is listed in the manifest but absent from the Asset folder, and its details are not available in our database, \ourtool{} conducts a thorough scan to check for any vulnerabilities, ensuring comprehensive coverage even for altered or inherently insecure libraries.

\begin{figure}
    \centering
    \begin{lstlisting}[
    language=ReportLog,
    breaklines=true,           %
    breakatwhitespace=true,    %
    columns=fullflexible       %
]{language=ReportLog, label={fig:scan_report}}
============================================================
              METASCANNER ANALYSIS REPORT
============================================================
Target World: @Experiment_Our_Malicious_World_v1@
Date: XXXX-XX-XX XX:XX:XX
Scan Modules: Object, Script, Library, QR_Code
============================================================
[MODULE 1] OBJECT SCANNER 
> Policy: 'Transparency Check (Alpha == 0)'
  - DETECTED: 'Intercept_Key' (MeshRenderer)
  - DETECTED: 'DOS_QR' (MeshRenderer)
  - DETECTED: 'Guard' (MeshRenderer)
  - DETECTED: 'VRCMirror' (MeshRenderer)
  
> Policy: 'Spatial Overlap (Position/Scale Match)'
  - DETECTED: 'Clickjacking_Button', 'Normal_Button'
  - DETECTED: 'keypad01', 'keypad02'
  => Potential 'Layered Click Redirection Attack'
  
[MODULE 2] SCRIPT SCANNER
> Suspicious API Calls Detected:
  - WARNING: Assets/Scripts/Clickjacking.cs
    -> Method: Application.OpenURL()
    -> Context: Potential forced redirection
	- WARNING: Assets/Scripts/TCP_Client.cs
    -> Method: System.Net.Sockets.Socket
    -> Context: External network communication attempt
    
[MODULE 3] QR SCANNER
> Decoded QR Codes from Assets:
  - Image: Assets/Prefabs/W119dJ.jpg
    -> URL: https://m.site.naver.com/119dJ
    -> Status: SUSPICIOUS (Shortened URL)
  
[MODULE 4] Library SCANNER
> Poisoned Library Detected:
  - WARNING: com.unity.nuget.newtonsoft-json@1.1.2
    -> Status: HASH MISMATCH
		
============================================================
FINAL SUMMARY:
- Total Suspicious Objects: 5
- Total Risky Scripts: 2
- Total Malicious QR Codes Image: 1
- Total Poisoned Library: 1
============================================================
    \end{lstlisting}
    \caption{An example of a security analysis report generated by \ourtool{} showing detection of objects, scripts, libraries, and QR codes.}
    \label{fig:MetaScanner_Report}
\end{figure}

\begin{table}
    \centering
    \caption{List of 38 worlds collected from the Unity Asset Store, detailing the number of objects and components in each world.}
    \resizebox{\linewidth}{!}{%
    \begin{tabular}{rlrr}
        \toprule
        \textbf{Index} & \textbf{World Name} & \textbf{\# Objects} & \textbf{\# Components} \\
        \midrule
        0 & Fantasy World & 4,318 & 12,941 \\
        1 & Toon Environments - World Creator Pack Lite & 260 & 730 \\
        2 & Little Low Poly World - LITE SRP/URP & 244 & 958 \\
        3 & Free Cartoon world & 119 & 449 \\
        4 & Free Demo Of Low Poly Space Alien Worlds 3D Asset Pack & 39 & 150 \\
        5 & Low poly alien world & 736 & 2,656 \\
        6 & World in the Pocket - Free & 21 & 75 \\
        7 & Unity Terrain - URP Demo Scene & 362 & 925 \\
        8 & Mountain Terrain, Rocks and Tree & 30 & 116 \\
        9 & Fantasy landscape & 364 & 1,209 \\
        10 & House Interior - Free & 339 & 1,085 \\
        11 & Office Pack - Free & 475 & 1,632 \\
        12 & 3D Realistic Terrain Free & 14 & 35 \\
        13 & Military FREE - Low Poly Asset Pack by ithappy & 31 & 827 \\
        14 & Real New York City Vol. 1 & 21,280 & 47,837 \\
        15 & Lowpoly Nature Bundle & 104 & 310 \\
        16 & Free Open Building & 1,028 & 3,347 \\
        17 & Fantasy House Bundle & 42 & 116 \\
        18 & Stylized Pirate Ship & 52 & 198 \\
        19 & POLY - Lite Survival Collection & 265 & 918 \\
        20 & Low-Poly Medieval Market & 2,574 & 9,175 \\
        21 & Furniture Free - Low Poly Asset Pack by ithappy & 44 & 168 \\
        22 & All In One - Heightmaps & 71 & 177 \\
        23 & Fristy Fantasy Lands 1 & 13 & 31 \\
        24 & Hill Rock Mountain Terrain & 42 & 161 \\
        25 & Cartoon Race Track - Oval & 327 & 1,002 \\
        26 & Fantasy FREE - Low Poly Asset Pack by ithappy & 56 & 186 \\
        27 & Wooden House - Free - Low Poly & 41 & 105 \\
        28 & Lite Farm Pack - Low Poly 3D Art by Gridness & 47 & 180 \\
        29 & The village well & 4 & 13 \\
        30 & Winter Mountains and Stamps & 39 & 118 \\
        31 & Small Town America - Streets - FREE & 167 & 500 \\
        32 & Hills, Mountains and Lakes - Terrains, Stamps and Brushes & 10 & 26 \\
        33 & Lowpoly Cactus Pack & 16 & 46 \\
        34 & Lowpoly Chest Monster & 41 & 59 \\
        35 & Flora-Form & 41 & 123 \\
        36 & Real New York City Vol. 2 & 32,660 & 74,283 \\
        37 & Low Poly Cartoon Mini Pack Free & 210 & 829 \\
        \bottomrule
    \end{tabular}
    }
    
    \label{tab:worlds_data}
\end{table}

\begin{figure}
    \centering
    \subfloat[World sizes]{
        \includegraphics[width=0.46\linewidth]{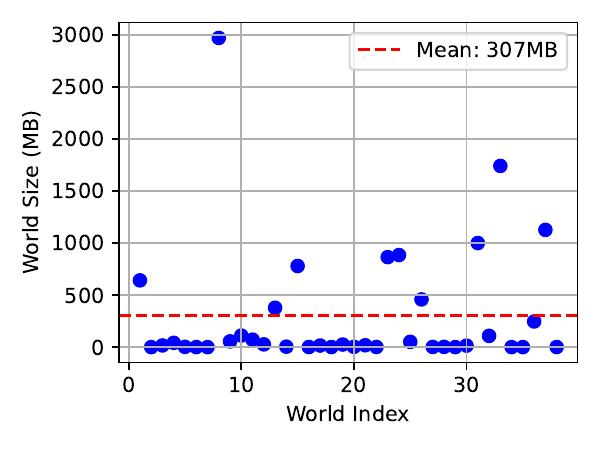}
        \label{fig:6-0}
    }
    \hfill
    \subfloat[Matched objects]{
        \includegraphics[width=0.46\linewidth]{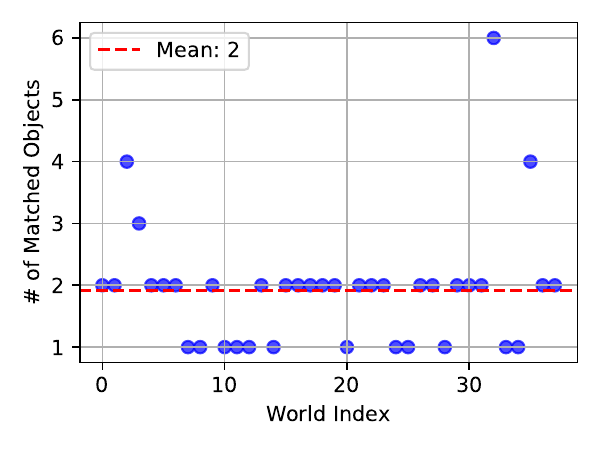}
        \label{fig:6-1}
    }
    \vspace{-0.1in}
    \subfloat[Matched components]{
        \includegraphics[width=0.46\linewidth]{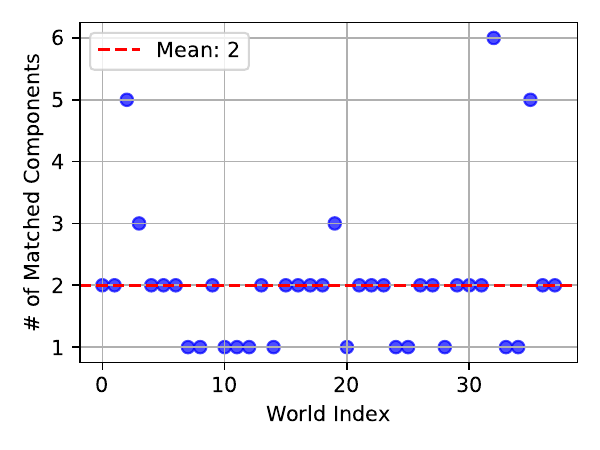}
        \label{fig:6-2}
    }
    \hfill
    \subfloat[Matched scripts]{
        \includegraphics[width=0.46\linewidth]{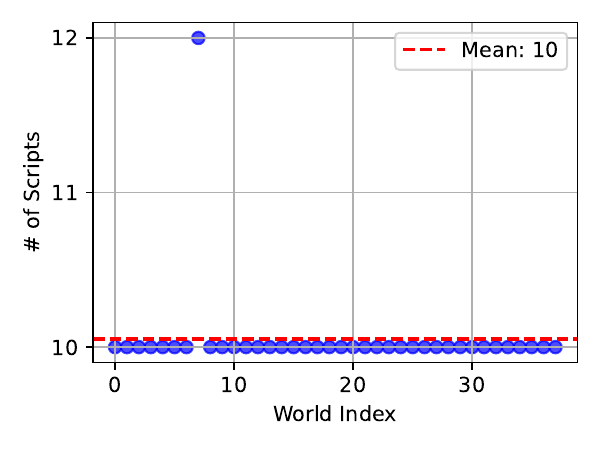}
        \label{fig:6-3}
    }
    \caption{Distribution of world sizes, matched objects, components, and scripts for 38 collected worlds.}
    \label{fig:6}
\end{figure}

\subsection{Dataset}
\label{subsec:dataset}

To evaluate the applicability of \ourtool{}, we aimed to inspect worlds published in VRChat. However, since VRChat World packages are not publicly accessible, we used the Unity Asset Store~\cite{unity_asset_store}. The Asset Store provides a wide range of Unity assets uploaded by various users. We collected 38 assets that are (i) free, (ii) tagged as 3D, and (iii) resemble the virtual worlds found on social VR platforms. These assets vary in size and contain numerous objects, components, and scripts. Figure~\ref{fig:6-0} illustrates the distribution of collected world sizes, and Table~\ref{tab:worlds_data} shows the list of worlds and their numbers of objects and components.

\subsection{Security Evaluation}

\begin{table*}
\centering
\caption{Summary of detection results for the proposed attacks by \ourtool{}.}
\scriptsize
\begin{tabular}{l L{3cm} L{1.5cm} L{2cm} L{5.0cm}}
\toprule
\textbf{Attack Scenario} & \textbf{Deceptive Element (Target)} & \textbf{Detection Result} & \textbf{Responsible Module} & \textbf{Detection Detail} \\
\midrule
Layered Click Redirection &
Overlapped \& Transparent Buttons &
Detected &
Object Scanner &
Identified fully transparent objects (alpha = 0) that overlap or are nearly overlapping ($\leq 0.2$ on x, y, z axes) \\ \midrule

Raycasting Phishing &
Invisible Walls (Blocking Ray) &
Detected &
Object Scanner &
Flagged GameObjects with invisible meshes or abnormal transparency interfering with raycasting \\ \midrule

In-World Input Hijack &
Transparent Keypad \& Input Exfiltration &
Detected &
Script Scanner \& Object Scanner &
Identified overlapping keypad objects intercepting input, with additional alerts from suspicious API calls (e.g., \texttt{OpenURL} or socket usage) \\ \midrule

Avatar Quishing &
Malicious QR Code on Avatar &
Detected &
QR Scanner &
Decoded embedded QR codes and flagged suspicious URLs from image assets \\

\bottomrule
\end{tabular}
\label{table:PoC_Attack}
\end{table*}

We first evaluated whether \ourtool{} can detect our proof-of-concept (PoC) attacks implemented in Section~\ref{section:user_study}. Table~\ref{table:PoC_Attack} summarizes the detection results of our defense across the four scenarios. For each scenario, we analyzed the deceptive element involved, whether the attack was detected, and which component of \ourtool{} identified the behavior during evaluation.

\vspace{0.05in}

\noindent\textbf{Layered Click Redirection.} As mentioned in Section~\ref{subsec:e1_clickjacking}, this attack relies on overlapped and transparent UI elements to mislead user interactions. Our Object Scanner first examines the alpha value of UI elements. However, since transparent objects are frequently used legitimately in Unity environments, transparency alone is insufficient for reliable detection. To address this, the scanner additionally analyzes the spatial position of objects. Specifically, it detects not only fully overlapping objects but also objects placed within a small positional offset to evade exact overlap checks. Objects whose relative positions differ by less than 0.2 units along each of the x, y, and z axes are treated as overlapping and flagged accordingly. Through evaluation, we confirmed that the UI elements of button position become perceptually distinguishable as separate buttons when their positional difference exceeds 0.2, justifying this threshold.

\vspace{0.05in}

\noindent\textbf{Raycasting Phishing.} This attack uses in-world objects to intercept or manipulate raycasting-based interactions (Section~\ref{subsec:e2_raycasting}). While such attacks may involve transparency or invisible meshes, successful interception also requires these objects to be spatially aligned with legitimate interaction targets. Accordingly, this scenario was detected primarily by the Object Scanner through its overlapping object detection policy, which identified objects positioned to intercept raycasting intended for legitimate UI or interaction elements.

\vspace{0.05in}

\noindent\textbf{In-World Input Hijack.} In this attack, a transparent keypad was placed on top of an existing legitimate keypad to capture user input (Section~\ref{subsec:e3_in-world}). Because this attack fundamentally depends on spatial overlap between the deceptive and legitimate input elements, it was first detected by the Object Scanner's overlapping object analysis. In addition, when the collected input was subsequently transmitted to an external server, the Script Scanner flagged suspicious API usage, including external URL access and socket-based communication, providing a secondary detection signal.

\vspace{0.05in}

\noindent\textbf{Avatar Quishing.} This attack involves embedding malicious QR codes within avatar assets (Section~\ref{subsec:e4_quishing}). Our QR Scanner inspects image files contained in the asset package and attempts to decode them as QR codes. While decoding arbitrary benign images typically yields meaningless or invalid results, images containing QR codes consistently decode into well-formed URLs. The extracted URLs are then analyzed using a URL inspection API, which successfully identified links leading to malicious destinations. Additionally, the system supports a whitelist-based policy that restricts allowed QR code destinations to a predefined set of trusted domains.

Overall, the results demonstrate that all four proposed attack scenarios were successfully detected during evaluation. Each scenario was flagged based on observable implementation characteristics, with different defense components contributing complementary detection signals. Figure~\ref{fig:MetaScanner_Report} presents the scan results of a world where our proposed attack was implemented.

\subsection{Detection Accuracy}
\label{subsec:detection_accuracy}

We next evaluated whether \ourtool{} can detect attacks deployed beyond our PoC environment using the 38 real-world environments collected from the Unity Asset Store (Section~\ref{subsec:dataset}). %
Three of the four attacks --- Layered Click Redirection, Raycasting Phishing, and In-World Input Hijack --- are implemented as world content. We therefore deployed each attack in 30 of the 38 worlds, excluding eight whose layouts did not provide a plausible location for the required UI elements. Each modified world was configured to allow users to approach and interact with the attacks as they would in a live virtual world. For each world-based attack, we included three differently configured variants in every world. For Avatar Quishing, we evaluated 10 distinct avatar assets using two different deployment methods.

We measured both the number of attack instances detected by \ourtool{} and false positives (FPs) among benign components originally present in each world, excluding objects introduced by the attacks. For overlap-based detection, the Object Scanner determines whether two objects coincide using the intersection over union (IoU) of their bounding volumes. We use $\mathrm{IoU} \ge 0.95$ as the default, allowing a $5\%$ tolerance from complete overlap, and additionally evaluate threshold values of $1.00$ and $0.80$ to examine sensitivity to this parameter.

\begin{figure}
    \centering
    \subfloat[Layered click redirection]{
        \includegraphics[width=0.46\linewidth]{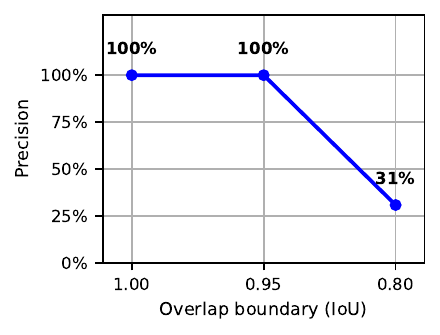}
        \label{fig:precision_overlap}
    }
    \hfill
    \subfloat[Raycasting phishing]{
        \includegraphics[width=0.46\linewidth]{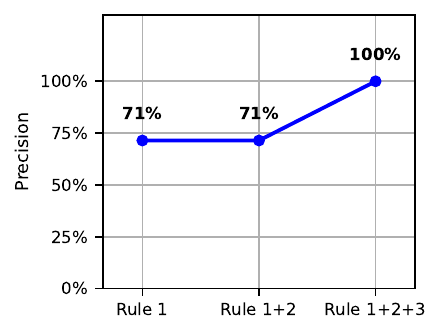}
        \label{fig:precision_raycast}
    }
    \caption{Precision under different detection-policy settings. (a) For Layered Click Redirection, lowering the overlap threshold to IoU = 0.80 introduces 67 false positives, reducing precision to 31\%. (b) For Raycasting Phishing, adding the QR-code condition (Rule 3) eliminates the 12 false positives produced by Rules 1–2, increasing precision to 100\%.}
    \label{fig:precision}
\end{figure}

\noindent\textbf{Layered Click Redirection.} We evaluated the clickjacking object at three degrees of overlap with the decoy object: $\mathrm{IoU}=1.00$ (complete overlap), $0.95$, and $0.80$. \ourtool{} detects all 30 instances at each setting. At complete overlap and at $0.95$, no false positives are observed. 
At $0.80$, however, 67 benign components across four worlds were flagged, reducing precision to $31\%$ (Table~\ref{tab:attack_detection} and Figure~\ref{fig:precision_overlap}).
These false positives were primarily decorative environmental elements placed in dense clusters, such as grass, bricks, and mushrooms. Thus, the relaxed boundary substantially increases the review burden, which we recognize as a limitation of using a lower overlap threshold.

\begin{table}
\centering
\caption{Detection results for the three world-based attacks across 30 virtual worlds and Avatar Quishing across 10 avatars. Detected indicates the number of attack instances identified by MetaScanner, while FP indicates the number of benign components incorrectly flagged, excluding components introduced as part of the attacks.}
\label{tab:attack_detection}
\small
\setlength{\tabcolsep}{4pt}
\begin{tabular}{llcc}
\toprule
\textbf{Attack} & \textbf{Setting} & \textbf{Detected} & \textbf{FP} \\
\midrule
\multicolumn{4}{l}{\emph{World-based attacks} ($n=30$ worlds)}\\
\midrule
Layered Click   & $\mathrm{IoU}=1.00$        & 30/30 & 0  \\
Redirection     & $\mathrm{IoU}=0.95$     & 30/30 & 0  \\
                & $\mathrm{IoU}=0.80$     & 30/30 & 67 \\
\midrule
In-World Input  & $\mathrm{IoU}=1.00$        & 30/30 & 0  \\
Hijack          & $\mathrm{IoU}=0.95$     & 30/30 & 0  \\
                & $\mathrm{IoU}=0.80$     & 30/30 & 67 \\
                & mismatched shape        & 0/30  & 0\\
\midrule
Raycasting      & rules 1--2 (object only)& 30/30 & 12 \\
Phishing        & rules 1--3 (with QR)    & 30/30 & 0   \\
\midrule
\multicolumn{4}{l}{\emph{Avatar-based attack} ($n=10$ avatars)}\\
\midrule
Avatar Quishing          & embedded QR             & 10/10 & 0   \\
        & fragmented QR           & 0/10 & 0  \\
\bottomrule
\end{tabular}
\end{table}

Nevertheless, relaxing the boundary does not provide a practical advantage to the adversary. At approximately $0.80$ IoU, the interaction highlight rendered around interactable objects, such as the outline shown in VRChat, makes the misaligned second object visually apparent to the user. An adversary that reduces the overlap enough to evade the stricter boundary therefore sacrifices the visual concealment required for the attack. 
These results support $\mathrm{IoU}=0.95$ as the default operating point.

\noindent\textbf{In-World Input Hijack.}
We placed a transparent capturing keypad over the legitimate one in all 30 worlds and evaluated IoU values of $1.00$, $0.95$, and $0.80$. As shown in Table~\ref{tab:attack_detection}, \ourtool{} detects every instance at all three settings, with false positives occurring only at $0.80$. At this level of overlap, however, the attack itself became ineffective because individual keys were no longer sufficiently aligned for the overlay to reliably capture the victim’s intended input.
Thus, relaxing the boundary results in a non-functional attack accompanied by false positives rather than successful evasion. Together with the click redirection results, these findings constrain the operating point from both sides. Both attacks require near-complete coincidence to function and are reliably detected at that degree of coincidence, while reducing the overlap degrades the attack itself. Thus, retreating below the $0.95$ boundary provides no practical advantage to the adversary.

We also examined limitations of overlap-based detection. A capturing interface whose shape differs substantially from that of the legitimate interface can fall outside the IoU-based boundary and was not detected in our mismatched-shape evaluation (no detection with the mismatched shape setting in Table~\ref{tab:attack_detection}). In addition, \ourtool{} analyzes creator-supplied interfaces contained in the uploaded world package. Client-owned interfaces created dynamically at runtime, such as platform settings or world-navigation menus, do not have corresponding world objects and escape \ourtool{} by construction.

\noindent\textbf{Raycasting Phishing.} 
The attack combines a transparent object that blocks raycasting with a QR-code notice that delivers the phishing payload. We evaluate three detection rules: Rule 1 identifies transparent objects by examining their alpha values and rendering configuration. Rule 2 further checks for a Mesh Collider and Rigidbody, indicating that the transparent object can interfere with user interaction. Applying Rules 1–2 detects all 30 attack instances but also flags 12 benign transparent objects, resulting in $71\%$ precision (Table~\ref{tab:attack_detection} and Figure~\ref{fig:precision_raycast}). These false positives primarily originated from benign invisible walls.

Rule 3 further searches for a QR code associated with the target object (or elsewhere) in the world to extract its embedded URL. Adding this condition eliminates all 12 false positives while retaining detection of all 30 attacks, increasing precision to $100\%$. Thus, the QR-code condition serves as an additional signal that distinguishes our phishing scenario from benign transparent colliders. An attack variant that blocks raycasting without a QR-code payload can still be flagged by Rules 1–2, but with lower specificity. If similar behavior is instead implemented through malicious scripts, the Script Scanner can provide an additional detection signal.

\noindent\textbf{Avatar Quishing.} 
We evaluated \ourtool{} using 10 distinct avatar projects downloaded from VRCMods~\cite{vrcmods}, each modified to contain a malicious QR code. \ourtool{} detects all 10 avatars containing directly embedded QR codes and produces no false positives on benign textures. However, static QR analysis is vulnerable to evasions in which no individual asset contains a complete, decodable QR code. In particular, fragmenting a QR code across multiple textures bypasses detection in all 10 evaluated avatars. Similarly, viewpoint-dependent compositions that form a valid QR code only when multiple objects are visually aligned can fall outside per-asset static analysis.

Note that \ourtool{} is evaluated only for extracting URLs rather than determining whether the URLs themselves are malicious. Social VR platforms can further apply their own allowlists, blocklists, or reputation services (e.g., Google Safe Browsing) to assess the extracted URLs. This design keeps the evaluation of \ourtool{} for Avatar Quishing independent of any particular external threat-intelligence service, while the final URL classification depends on the platform’s downstream security policy.

\subsection{Scanning Effectiveness}
\label{subsec:scan_effect}

\begin{figure*}[t]
    \centering
    \subfloat[Total scanning]{
        \includegraphics[width=0.24\textwidth]{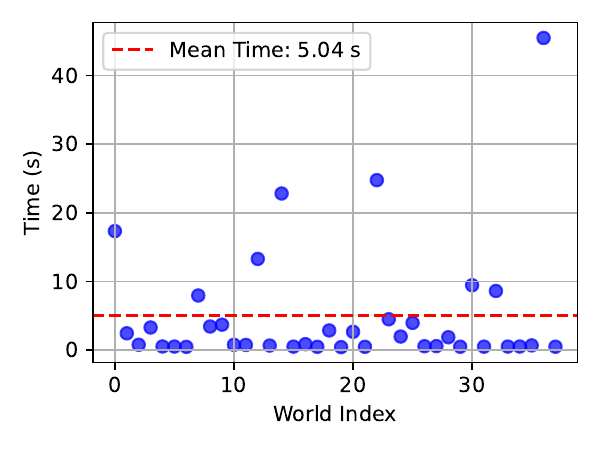}
        \label{fig:5-1}
    }
    \subfloat[Library scanning]{
        \includegraphics[width=0.24\textwidth]{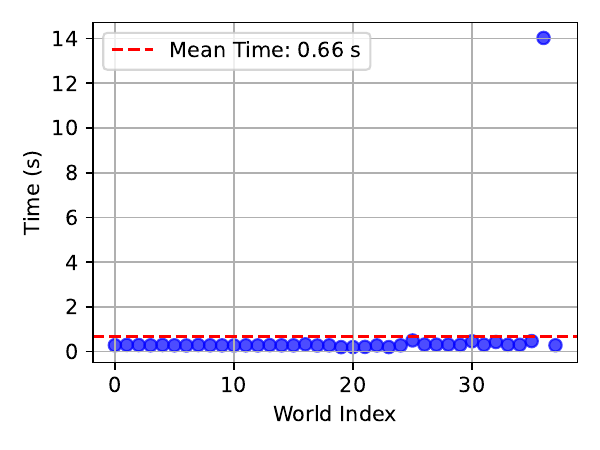}
        \label{fig:5-2}
    }
    \subfloat[Object scanning]{
        \includegraphics[width=0.24\textwidth]{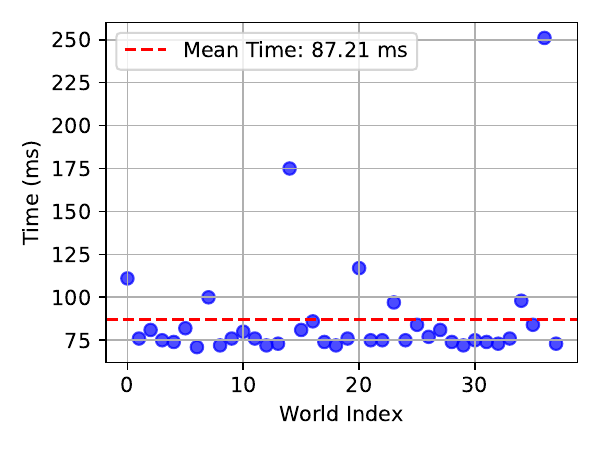}
        \label{fig:5-3}
    }

    \subfloat[Script scanning]{
        \includegraphics[width=0.24\textwidth]{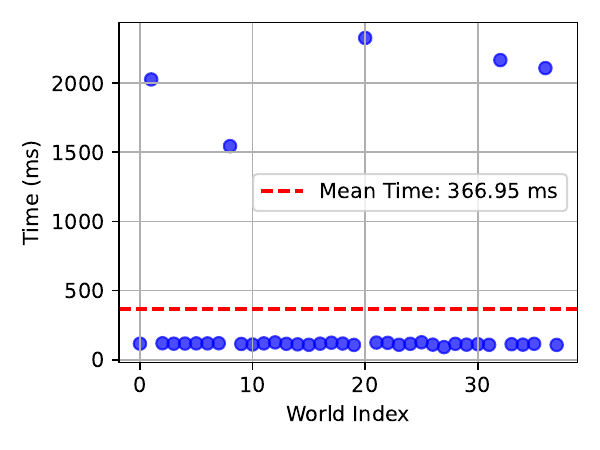}
        \label{fig:5-4}
    }
    \subfloat[QR code scanning]{
        \includegraphics[width=0.24\textwidth]{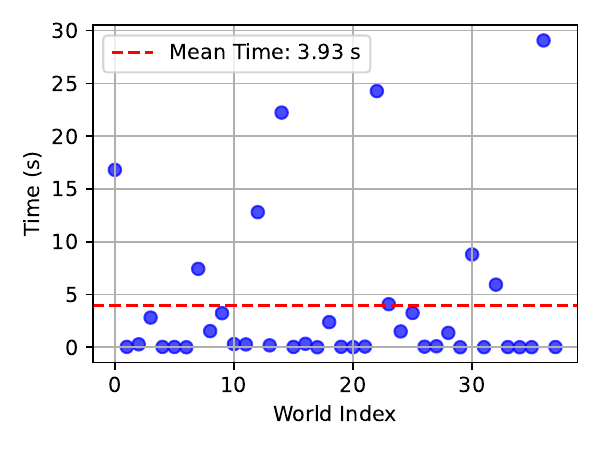}
        \label{fig:5-5}
    }
    \caption{Scanning time distribution across 38 worlds for total, library, object, script, and QR code scanning.}
    \label{fig:5}
\end{figure*}

We evaluated \ourtool{} to verify its ability to detect objects and scripts within a virtual world based on defined policies. For object and component scanning, we used a dummy policy to confirm the process, checking settings of automatically created objects like MainCamera and Light when a Unity scene initializes. We also applied three security policies to detect suspicious objects: (i) overlapping objects using the same scale and prefabs within a proximity of \textbf{n=3}, (ii) transparent objects with a material alpha value of 0, and (iii) trigger objects containing a \texttt{Rigidbody} component with the \texttt{isTrigger} property enabled, commonly used to trigger actions in Unity. For script scanning, we targeted simple but potentially vulnerable functions, focusing on those attempting external connections, such as \texttt{OpenURL} and \texttt{Socket}. Unlike object policies, which require more complex evaluations, script policies rely on straightforward string comparisons, making them easily adjustable.

Figures~\ref{fig:6-1} and \ref{fig:6-2} illustrate the number of matched objects and components across various worlds. On average, only 6\% of all objects and 2\% of all components matched our policies. This suggests that suspicious objects, such as transparent or overlapping objects used in our UI attacks, are relatively rare in most worlds; thus, \ourtool{} can detect such objects if adversaries exploit them. Specifically, only two basic objects were detected across the worlds in our results, such as the Camera and the Light object, which are default objects in the Unity scene, and no suspicious objects were found. 
A noticeable world is ``Hills, Mountains and Lakes - Terrains, Stamps and Brushes,'' which showed more matched objects (i.e., 6). This occurred because the asset contained multiple scenes, and three of these scenes were scanned simultaneously. Each scene had two dummy policies applied, resulting in a total of six detected objects. It is important to note that in this case, no genuinely suspicious objects were identified.

Figure~\ref{fig:6-3} shows that \ourtool{} detected 10 suspicious scripts on average in most worlds. These scripts were mostly identified in libraries commonly used for collaboration in Unity, such as the Collaborate library, the Visual Studio library, a popular IDE, and the Timeline library, which supports the creation of cutscenes and cinematic sequences. It was found that these scripts utilized functions like \texttt{OpenURL} and \texttt{Socket}. While this functionality in itself is not problematic, there is a potential for misuse. Specifically, if a social VR platform merely restricts user script compilation for certain functions, adversaries could exploit functions from these packages to circumvent such restrictions. Therefore, careful attention is required to prevent such misuse. 

\subsection{Scanning Time}

We measured \ourtool{}'s processing time for each world scan. As shown in Figure~\ref{fig:5-1}, the total scanning time averaged 5 seconds, indicating reasonable performance for a proactive solution. An exception is the world ``Real New York City Vol. 2,'' which took 45 seconds due to its large number of objects (32,660) and components (74,283). Of this, approximately 29 seconds were spent on QR code scanning, largely because over 225 image files were embedded in object materials. Scanning libraries, objects, and scripts averaged less than 1 second (Figures~\ref{fig:5-2}–\ref{fig:5-4}), accounting for 13.1\%, 1.7\%, and 7.3\% of the total time, respectively. The main bottleneck was QR code scanning, which took 3.9 seconds on average (Figure~\ref{fig:5-5}) due to image object analysis.

\section{Discussion}

\begin{table*}
\centering
\caption{UI attacks and strategies implementable on each social VR platform. (\CIRCLE: implementable, \Circle: not implementable, $\ast$: external HTTPS requests partially supported)}

\scriptsize
\begin{tabular}{ l l l @{} c c c c}
\toprule
\multirow{2}{*}{\textbf{Scenario}} & \multirow{2}{*}{\textbf{Strategy}} & \multirow{2}{*}{\textbf{Need}} & \multicolumn{4}{c}{\textbf{Social VR Platform}} \\ \cmidrule{4-7}

&  &  & VRChat~\cite{vrchat} & Spatial~\cite{spatial_homepage} & Decentraland~\cite{decentraland_homepage} & Rec Room~\cite{recroom} \\

\midrule
\multirow{2}{*}{Layered Click Redirection} & Same Size & No Flickering & \CIRCLE & \CIRCLE$\ast$ & \CIRCLE & \CIRCLE \\
                                     & Invisible Object & Transparent & \CIRCLE & \CIRCLE$\ast$ & \CIRCLE & \CIRCLE \\
\midrule
Raycasting Phishing & Block Raycasting & Transparent & \CIRCLE & \CIRCLE & \CIRCLE & \CIRCLE \\
\midrule
\multirow{2}{*}{In-World Input Hijack} & Intercept Raycasting & Transparent \& Raycasting & \CIRCLE & \Circle & \CIRCLE & \CIRCLE \\
                                      & Intercept Key Input & Transparent \& Input Replica & \CIRCLE & \CIRCLE$\ast$ & \CIRCLE & \CIRCLE \\
\midrule
Avatar Quishing & - & Custom Avatar Upload & \CIRCLE & \CIRCLE & \CIRCLE & \Circle \\
\bottomrule
\end{tabular}
\label{table:platform_support}
\end{table*}

\noindent\textbf{Generalizability of UI Attacks.} We also examined the feasibility of our attacks across other social VR platforms, including Rec Room~\cite{recroom}, Spatial~\cite{spatial_homepage}, and Decentraland~\cite{decentraland_homepage}, as summarized in Table~\ref{table:platform_support}. Overall, most of the attacks are implementable, with a few exceptions. For example, Rec Room does not fully support custom avatar creation—only verified users are permitted to upload content. Consequently, we were unable to implement the avatar quishing attack, as it prevents the upload of QR codes. This limitation may be relaxed in the future as avatar customization features continue to evolve. Additionally, Spatial uses circular interaction zones around objects instead of raycasting. This mismatch causes raycasting to inaccurately detect interaction targets. To overcome this, attackers can place an invisible replica of the keypad at the same position as the original, leveraging platform-provided input features to capture controller inputs precisely. Finally, some attacks may become even more impactful depending on the platform’s specific capabilities. In particular, Spatial partially supports external HTTPS requests, which could allow adversaries to transmit data to malicious servers or redirect users to harmful websites.
 
However, the four attacks in this study are not intended to constitute an exhaustive taxonomy of deceptive UI attacks in social VR. Rather, they represent concrete attack patterns derived from selected interaction primitives exposed by the VRChat and Unity engine SDK. Other platforms, interaction mechanisms, and runtime behaviors may enable additional attacks beyond those considered in this paper. Systematically characterizing this broader attack space remains an important direction for future work.

\vspace{0.05in}
\noindent\textbf{Comparison to Existing Safety Systems.} 
Social VR platforms often provide an in-house safety measure to protect their users against malice. For example, VRChat comes with the VRChat safety and trust system~\cite{vrchat_safety}: a user can configure what actions (e.g., hearing voices or displaying custom avatar animations) are allowed for other users based on their predefined trust level. Additionally, VRChat leverages Easy Anti-Cheat (EAC)~\cite{vrchat_eac} to block all modified clients, which can act as malicious clients attempting to exploit vulnerabilities in the platform.
The key difference of such systems is that the primary goal is to protect against potential nuisance activities from \emph{other participating users} in a virtual world. In contrast, \ourtool{} addresses UI attacks introduced by content creators, who have greater control over virtual environments than individual participants. By operating at the upload stage rather than runtime, \ourtool{} proactively inspects creator-supplied content before publication. It detects structural threats in virtual world packages, including transparent or overlapped objects, suspicious API calls, embedded QR URLs, and modified library hashes, while not covering purely runtime logic, social engineering, or voice-based deception. \ourtool{} can be integrated into the build and upload pipeline, such as the VRChat SDK workflow, to inspect world or avatar files before publication.

\vspace{0.05in}
\noindent\textbf{Perception of Harmful Content in Social VR.}
In all our experiments, we ensured that all videos and resources used were carefully selected to avoid any harmful content.
However, the impact of our proposed attacks could become much more severe when harmful content is played or used.
Discussions in the VRChat community have raised concerns about situations where a participant might turn on harmful video without consent, and even cases where it is unclear who turned on the video \cite{reddit_concern, reddit_discussion}.
Despite these concerns, the VRChat Community Guidelines~\cite{vrchat_community_guideline} simply state that harmful content should not be used, without providing specific regulations or clear definitions of the harmful content.
With the growing use of social VR, it is increasingly important to have an open discussion that includes user perceptions of VR interactions to establish a clear definition of harmful experiences.

\vspace{0.05in}

\noindent\textbf{False Positives in \ourtool{}. } \ourtool{} is inherently subject to false positives because visual properties, such as transparency and spatial overlapping, are widely employed features in standard Unity World development. In our evaluation across 38 collected worlds, we observed that the detection of overlapping objects is sensitive to the proximity threshold; specifically, depending on how the distance threshold is adjusted, legitimate objects placed in close proximity were occasionally misclassified as suspicious.

To mitigate this, \ourtool{} employs a sequential verification process rather than relying on object properties alone. First, the Object Scanner acts as a primary filter, identifying ``Suspicious Objects'' based on visual heuristics (e.g., invisible colliders blocking interaction). Subsequently, these flagged objects are analyzed by the Script Scanner to determine if they are associated with malicious logic. This secondary verification effectively minimizes false positives for benign experiential elements, such as mirror mazes, which utilize transparency for immersion but lack harmful scripts. By requiring this two-step confirmation, \ourtool{} ensures that legitimate design elements are not mistakenly flagged, allowing it to focus exclusively on detecting actual deceptive UI attacks.

Nevertheless, a limitation regarding false positives remains. Since \ourtool{} fundamentally classifies transparent objects as potential threats and specifically identifies overlapping objects as primary indicators of In-World Input Hijack and Layered Click Redirection attacks, legitimate objects that strictly adhere to these structural patterns may still be flagged as risks. Therefore, while our multi-step analysis significantly reduces noise, completely eliminating false alarms for complex benign structures remains a challenge. %

\begin{table*}
\centering
\caption{Comparison of previous studies and our work. (\CIRCLE: applicable, \Circle: not applicable, N/A: no attack is demonstrated) \textit{Platform-native}: the attack is demonstrated on a commercially deployed platform or application. \textit{Without Malware}: the attack requires neither installing a malicious application nor modifying the client application. \textit{Low Privilege}: the attacker needs no elevated access (e.g., developer mode, root, or network/server control). \textit{Social VR based}: the attack targets a multi-user social VR setting.}
\scriptsize
\begin{tabular}{l l l l l}
\toprule
\textbf{Study} & \textbf{Platform-native} & \textbf{Without Malware} & \textbf{Low Privilege} & \textbf{Social VR based} \\
\midrule
Lee et al.~\cite{lee2021adcube} & \Circle & \CIRCLE & \CIRCLE & \Circle \\
Tseng et al.~\cite{tseng2022dark} & \Circle & \Circle & \Circle & \Circle \\
Cheng et al.~\cite{cheng2024user} & \Circle & \Circle & \CIRCLE & \Circle \\
Cheng et al.~\cite{cheng2023exploring} & \Circle & \Circle & \CIRCLE & \Circle \\
Yang et al.~\cite{yang2024inception} & \CIRCLE & \Circle & \Circle & \CIRCLE \\
Ramirez et al.~\cite{ramirez2024deceptive} & N/A & N/A & N/A & \Circle \\
Valluripally et al.~\cite{valluripally2021modeling} & \Circle & \Circle & \Circle & \CIRCLE \\
Vondráček et al.~\cite{vondravcek2023rise} & \CIRCLE & \Circle & \CIRCLE & \CIRCLE \\
Sajid et al.~\cite{sajid2025just} & \Circle & \Circle & \CIRCLE & \Circle \\ \midrule
\textbf{Our work} & \CIRCLE & \CIRCLE & \CIRCLE & \CIRCLE \\
\bottomrule
\end{tabular}
\label{table:related_work}
\end{table*}

\section{Related Work}

\noindent\textbf{UI Attacks in AR/VR.} Research has exposed UI vulnerabilities in AR and VR that allow adversaries to deceive users. Lee et al.~\cite{lee2021adcube} demonstrated ad fraud in WebXR through deceptive clicks. Tseng et al.~\cite{tseng2022dark} showed how VR perception manipulation leads to unsafe behaviors. Cheng et al.~\cite{cheng2024user,cheng2023exploring} identified AR vulnerabilities like view manipulation and clickjacking, while Yang et al.~\cite{yang2024inception} introduced inception attacks that exploit VR debugger mode to alter UIs. Inspired by these, we propose new UI attacks targeting social VR platforms, affecting a larger user base with lasting impact.

\vspace{0.05in}
\noindent\textbf{Deceptive UI in Social VR.}
Deceptive UI manipulates users' perception and attention to induce decisions against their intent. In VR, such manipulations are even subtler and more effective through multisensory feedback and realistic simulations~\cite{hadan2024survey}. These findings support our study results, which show that manipulations in social VR can compromise user autonomy, collect personal data, and lead to psychological and economic harm. Although visible harm has not yet been widely reported in social VR, Ramirez et al.~\cite{ramirez2024deceptive} show that the FTC has already penalized companies for using deceptive interfaces, demonstrating the urgent need for research and regulation on deceptive UI in social VR.

\vspace{0.05in}
\noindent\textbf{Security Issues in Social VR Platforms.} Previous studies have examined various security vulnerabilities in social VR platforms. Valluripally et al.~\cite{valluripally2021modeling} identified several attack vectors in social VR learning environments, including unauthorized access and network fault injections that can induce cybersickness in users. Vondráček et al.~\cite{vondravcek2023rise} described the ``man-in-the-room'' attack, where an adversary hides within a virtual world to eavesdrop on private conversations by maliciously patching the Bigscreen application. Concurrently, Mengascini et al.~\cite{mengascini2026omniscience} showed that world creators can covertly surveil users through standard creation tools, but they focus on privacy violations and explicitly exclude the deceptive UI attacks that our work addresses. In contrast, our work focuses on exploiting UI features in social VR platforms to manipulate users into unknowingly performing malicious actions.

\vspace{0.05in}

\noindent\textbf{Comparison with Previous Studies.} Table~\ref{table:related_work} compares our work and related studies. While similar attacks have been proposed, some attacks require installing malicious applications or modified clients~\cite{tseng2022dark,vondravcek2023rise}, network-level control~\cite{valluripally2021modeling}, or enabled USB debugging mode~\cite{yang2024inception}. Similarly, Cheng et al.~\cite{cheng2023exploring} assume an intentionally manipulative MR application. Additionally, the work by Sajid et al.~\cite{sajid2025just} requires malicious code, as their threat model assumes an attacker compromises the MR application’s security or functionality through third-party integration code. Although their click redirection and occlusion attacks resemble some of our techniques, they are demonstrated in a research prototype of remote collaborative MR rather than a deployed social VR platform. In contrast, our study does not rely on such malware or malicious applications.
Furthermore, although Yang et al.~\cite{yang2024inception} and Vondráček et al.~\cite{vondravcek2023rise} target commercial social VR applications, their attacks require developer mode on the victim's headset or a patched client application, respectively. In contrast, our study is the first to demonstrate the abuse of a social VR environment within an actively operating platform using only the content creation capabilities available to normal users.

\section{Conclusion}

This paper presents novel UI attacks that deceive users into performing malicious actions on social VR platforms. We demonstrate how adversarial content creators can embed stealthy attacks within virtual worlds and avatars, with particular emphasis on a representative clickjacking attack. We provide a comprehensive analysis of the proposed attack scenarios and report findings from an IRB-approved user study involving 30 participants, which confirms the effectiveness of these attacks.

To counter these threats, we propose \ourtool{}, a practical defense system that proactively analyzes user-created content to identify suspicious objects, scripts, and assets before deployment. Our evaluation results show that \ourtool{} can efficiently analyze large volumes of objects and scripts within seconds using a set of well-defined detection policies, successfully identifying all proposed attack scenarios. We believe this work contributes to improving the security and trustworthiness of social VR platforms by highlighting emerging UI-based threats and providing a deployable mitigation approach.

\section*{Acknowledgments}
This work was partially supported by Institute of Information \& Communications Technology Planning \& Evaluation (IITP) [RS-2023-00215700, Trustworthy Metaverse: blockchain-enabled convergence research] and the National Research Foundation of Korea (NRF) (No.~RS-2024-00457937) grant funded by the Korea government (MSIT).

\bibliographystyle{unsrt}

\appendix
\section{User Study Questionnaire}
\label{appendix:questionnaire}

The following questionnaire was administered to participants after they completed the tasks in the virtual world. The survey was conducted via Google Forms.

\subsection*{Demographics}
\textit{Note: Demographic information, such as age and gender, was collected during the preliminary survey process after participants were informed and signed the consent form.}
\begin{enumerate}
    \item \textbf{Do you frequently use VR devices?}
    \begin{enumerate}
        \item Often Use
        \item Sometimes Use
        \item Neutral
        \item Rarely Use
        \item Never Used
    \end{enumerate}
\end{enumerate}

\subsection*{Experiment 1: Layered Click Redirection (The Tomato Room)}
\begin{center}
    \includegraphics[width=0.6\linewidth]{figures/click_impl1.pdf}
\end{center}
\textit{Context: A room with two buttons that were supposed to spawn tomatoes.}

\begin{enumerate}
    \setcounter{enumi}{1} %
    \item \textbf{Did you notice any difference between the first button and the second button before pressing them?}
    \begin{enumerate}
        \item I noticed a difference.
        \item I did not notice any difference.
        \item Other: \underline{\hspace{3cm}}
    \end{enumerate}

    \item \textbf{(If you answered `No' to Q2) What was the reason you did not notice a difference?}
    \begin{itemize}
        \item (Open-ended response)
    \end{itemize}

    \item \textbf{When pressing the second button, what did you expect?}
    \begin{itemize}
        \item (Open-ended response)
    \end{itemize}

    \item \textbf{How did you feel when the video played on the screen?}
    \begin{itemize}
        \item (Open-ended response)
    \end{itemize}
\end{enumerate}

\subsection*{Experiment 2: Raycasting Phishing (The Fruit Room)}
\begin{center}
    \begin{minipage}{0.48\linewidth}
        \centering
        \includegraphics[width=\linewidth]{figures/dos_impl1.pdf}
    \end{minipage}
    \hfill
    \begin{minipage}{0.48\linewidth}
        \centering
        \includegraphics[width=\linewidth]{figures/dos_impl2.pdf}
    \end{minipage}
\end{center}
\textit{Context: A room where pressing a red button caused subsequent interactions to fail, displaying a warning.}

\begin{enumerate}
    \setcounter{enumi}{5}
    \item \textbf{Did you notice any special environmental changes or visual differences immediately after pressing the red button (excluding the appearance of the red button itself)?}
    \begin{itemize}
        \item Yes
        \item No
    \end{itemize}

    \item \textbf{What did you think when the fruit buttons stopped working and the warning message appeared?}
    \begin{enumerate}
        \item I thought the device was malfunctioning.
        \item I thought there was a bug in the room/world.
        \item I thought it was an intended feature of the room.
        \item Other: \underline{\hspace{3cm}}
    \end{enumerate}

    \item \textbf{(If you selected `Intended feature' in Q7) How do you think this feature was implemented?}
    \begin{itemize}
        \item (Open-ended response)
    \end{itemize}

    \item \textbf{Did you consider scanning the QR code with your smartphone?}
    \begin{enumerate}
        \item Yes
        \item No
        \item Other: \underline{\hspace{3cm}}
    \end{enumerate}
\end{enumerate}

\subsection*{Experiment 3: In-World Input Hijack (The Keypad Room)}
\begin{center}
    \includegraphics[width=0.6\linewidth]{figures/mitm_impl2.pdf}
\end{center}
\textit{Context: A room requiring password input via a keypad.}

\begin{enumerate}
    \setcounter{enumi}{9}
    \item \textbf{Did you notice anything special or strange about the keypad before inputting your number?}
    \begin{itemize}
        \item Yes
        \item No
    \end{itemize}

    \item \textbf{How did you feel when you learned that the numbers you entered were transmitted to a different location (hidden room)?}
    \begin{itemize}
        \item (Open-ended response)
    \end{itemize}

    \item \textbf{Can you guess how the number redirection function is implemented?}
    \begin{itemize}
        \item (Open-ended response)
    \end{itemize}
\end{enumerate}

\subsection*{Experiment 4: Avatar Quishing (The QR Code Room)}
\begin{center}
    \includegraphics[width=0.6\linewidth]{figures/qshing_impl1.pdf}
\end{center}
\textit{Context: A room where users had to choose between two QR codes (A and B).}

\begin{enumerate}
    \setcounter{enumi}{12}
    \item \textbf{Did you notice any difference between the two QR codes? (Differences in the QR pattern itself do not count.)}
    \begin{enumerate}
        \item I noticed a difference.
        \item I did not notice any difference.
    \end{enumerate}

    \item \textbf{(If you answered `Yes' to Q13) Please describe how you perceived the difference.}
    \begin{itemize}
        \item (Open-ended response)
    \end{itemize}

    \item \textbf{Could you perceive that one of the two QR codes actually belonged to (was attached to) an avatar?}
    \begin{enumerate}
        \item Yes
        \item No
    \end{enumerate}

    \item \textbf{Have you used QR codes in your daily life? (e.g., smart attendance, shared bike/scooter rentals)}
    \begin{enumerate}
        \item Yes
        \item No
    \end{enumerate}

    \item \textbf{(If you answered `Yes' to Q16) Have you ever doubted the authenticity of a QR code when scanning it with your smartphone?}
    \begin{enumerate}
        \item Yes
        \item No
        \item Other: \underline{\hspace{3cm}}
    \end{enumerate}
\end{enumerate}

\subsection*{General Feedback}
\begin{enumerate}
    \setcounter{enumi}{17}
    \item \textbf{If you have any other comments regarding this experiment, please feel free to write them down.}
    \begin{itemize}
        \item (Open-ended response)
    \end{itemize}
\end{enumerate}

\section{QR Code Scanning in Social VR}
\label{appendix:qr}

To assess whether social VR users have any practical means of vetting a QR code before visiting the URL it encodes, we surveyed the QR-scanning approaches that are actually in use within the VRChat community and examined whether any of them perform a safety or reputation check on the destination URL.

VRChat does not ship a built-in, general-purpose QR-code reader. A native scanner has been a commonly requested feature for several years, tracked publicly on VRChat's official feedback board with over 100 votes~\cite{vrchat-feedback-qr}. Notably, one commenter on that thread argued that adding a native QR/URL reader could in fact increase risk, since an in-world image (and therefore the QR code it displays) can be swapped out via dynamic image loading at any time without requiring an avatar or world re-upload, weakening any accountability that re-upload review would otherwise provide~\cite{vrchat-feedback-qr}. This underscores that the community itself is aware of QR-based image swapping as an attack surface, even in the absence of the concrete exploitation scenario we describe in this paper.

In the absence of an official reader, users rely on third-party, community-built tools. We identified three representative examples that are commonly circulated and recommended within the VRChat community:

\begin{itemize}
 \item \textbf{VRQR}~\cite{frostion-vrqr}: a standalone desktop companion application that watches the user's VRChat screenshot folder, decodes any QR code found in a newly captured screenshot, and automatically launches the encoded URL in the user's default browser the moment it is detected.
 \item \textbf{VRCQR}~\cite{yeusepe-vrcqr}: a paid, VRCFury-compatible Unity package that embeds QR-scanning directly into the user's avatar via OSC, allowing the avatar itself to detect and act on QR codes encountered in-world.
 \item \textbf{QRKicker}~\cite{qrkicker-github}: an open-source utility that similarly monitors VRChat screenshots for QR codes and opens the resulting website automatically.
\end{itemize}
We summarize these tools in Table~\ref{tab:qr-scanners}. None of the three surveyed tools perform any URL reputation lookup, malicious-domain detection, or user confirmation step before acting on a decoded URL; each is designed to minimize friction by opening the link as soon as it is recognized, which is precisely the behavior an attacker can exploit.

\begin{table}[h]
\centering
\caption{QR-code scanning approaches available to VRChat users.}
\resizebox{\linewidth}{!}{%
\label{tab:qr-scanners}
\begin{tabular}{l l c c}
\toprule
\textbf{Approach} & \textbf{Provider} & \textbf{Safety Check} & \textbf{Platform-integrated} \\
\midrule
VRChat~\cite{vrchat} & VRChat & N/A & Yes \\
VRQR~\cite{frostion-vrqr} & Community & No & No (external app) \\
VRCQR~\cite{yeusepe-vrcqr} & Community & No & Partial (avatar asset) \\
QRKicker~\cite{qrkicker-github} & Community & No & No (external app) \\
Smartphone & Device OEM & Depends on device & No \\
\bottomrule
\end{tabular}}
\end{table}

\begin{figure}
    \centering
    \includegraphics[width=\linewidth]{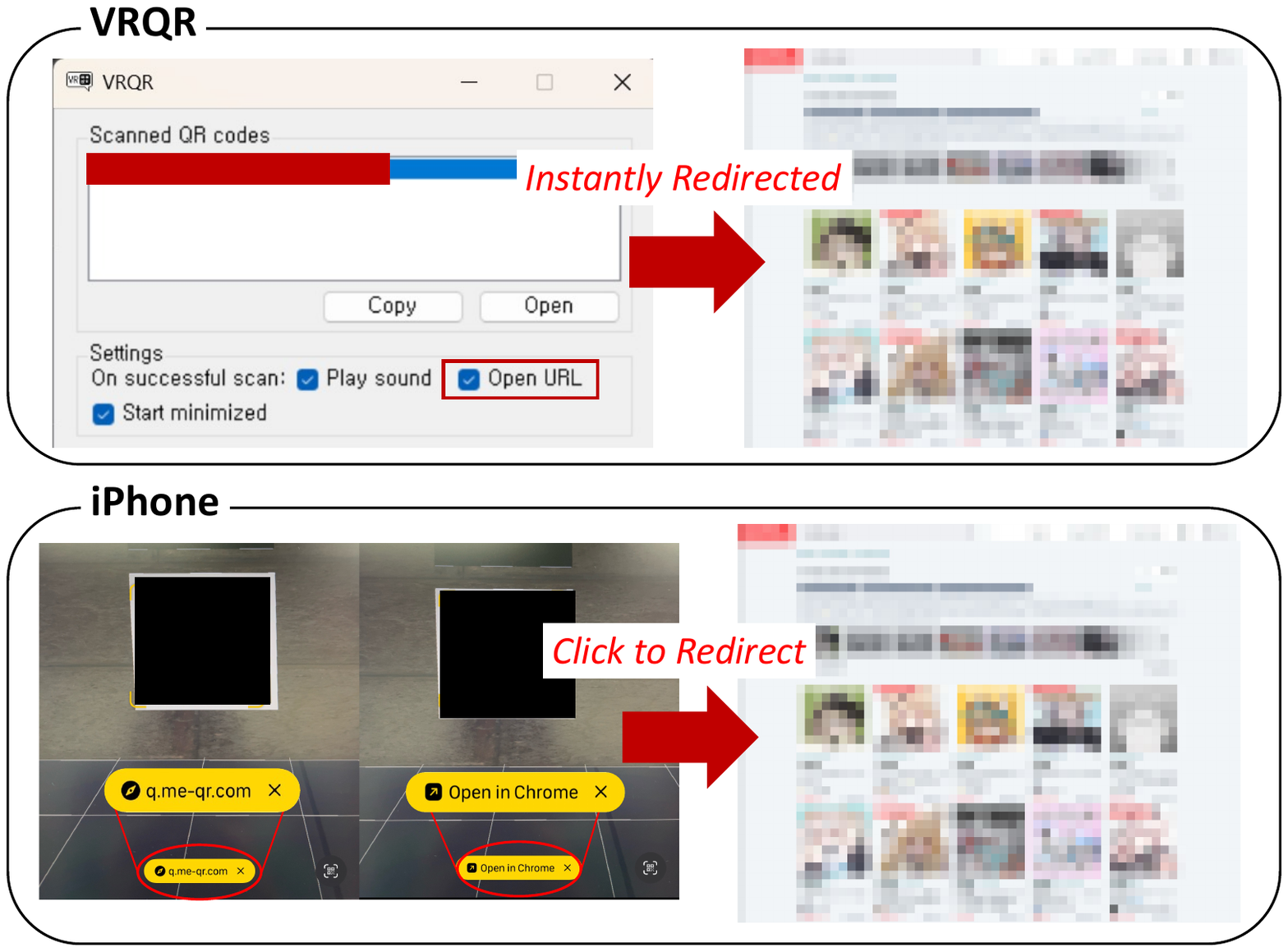}
    \caption{QR scanning interfaces in the tested settings. VRQR automatically opens the decoded URL with Open URL enabled by default (top). The iPhone Camera app displays the destination domain with Safari selected as the default browser and \texttt{Open in Chrome} with Chrome selected (bottom). Both iPhone settings require a tap to open the URL.}
    \label{fig:other_scanner}
\end{figure}

We further examined the scanning behavior of VRQR and the iPhone Camera app, as shown in Figure~\ref{fig:other_scanner}. In the version of VRQR we downloaded from its GitHub repository, the Open URL option was enabled by default. When a VRChat screenshot containing a QR code was captured, VRQR decoded the code and automatically opened the encoded URL in the default browser without requiring the user to select the URL or confirm that it should be opened. Although VRQR displayed the decoded result in its interface, opening the URL did not require prior inspection by the user.

For the iPhone Camera app, the information displayed before opening the URL differed depending on the default browser. With Safari selected as the default browser, the camera displayed the domain of the encoded URL. With Chrome selected, the camera displayed Open in Chrome without showing the destination URL on the button. In both tested settings, tapping the button opened the URL in the selected browser without an additional warning dialog. These observations illustrate differences in the information available before opening a QR destination. In our user study, participants used their own smartphones to access the questionnaire, so these examples do not establish the URL preview shown to each participant.

\bio{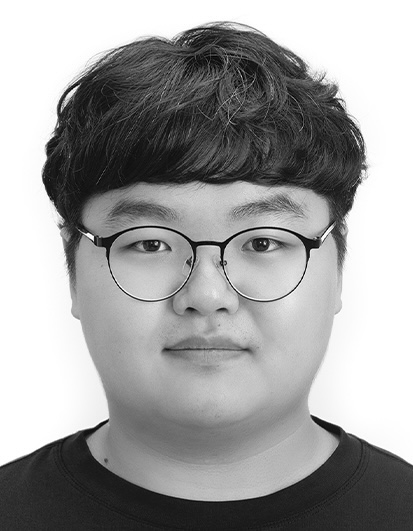}
\textbf{Junhee Lee} is currently a Research Associate at the CNS Lab (Computing and Network Security Lab) at Chungbuk National University. He received his B.S. in the School of Software Engineering and M.S. in Computer Science from Kwangwoon University. His research interests include security and ethical issues in VR/XR platforms.
\endbio

\vspace{0.5in}

\bio{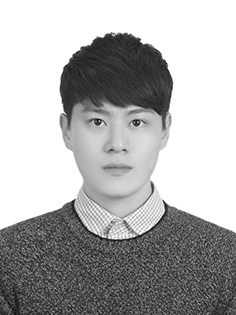}
\textbf{Seungwon Woo} is a Researcher at the Electronics and Telecommunications Research Institute (ETRI) and a Ph.D. student in the Graduate School of Information Security (GSIS) at KAIST. He received his M.S. degree in the Graduate School of Information Security from KAIST, and his B.S. degree in Computer Science and Engineering from Chungnam National University. His research interests lie in distributed systems and networking, with a focus on improving the security issues of AR/VR and decentralized systems.
\endbio

\vspace{0.2in}

\bio{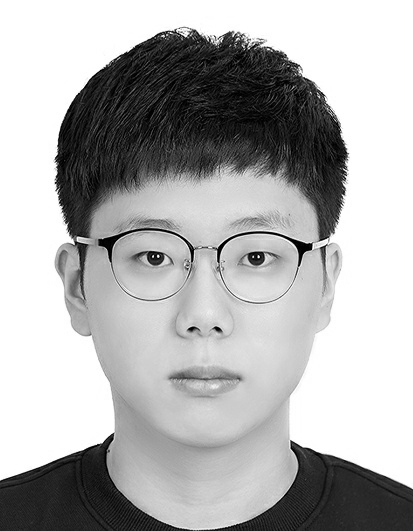}
\textbf{Minseok Kim} is an undergraduate research student at the CNS Lab (Computing and Network Security Lab), School of Software, Kwangwoon University. His research interests include security and ethical issues in metaverse environments, such as social VR platforms.
\endbio

\newpage

\bio{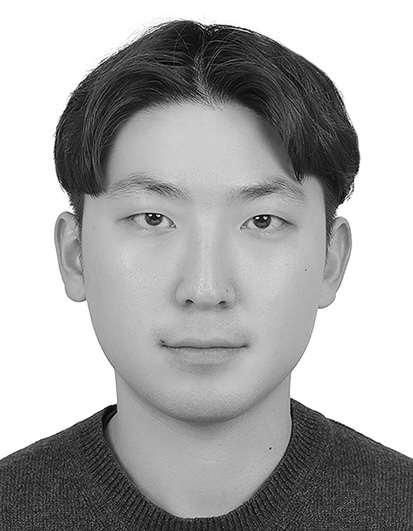}
\textbf{Jongseop Kim} is a master's student in Computer Science at Kwangwoon University. He received his B.S. in Mathematics with a minor in Software from Kwangwoon University. He is currently conducting research as a graduate student at the CNS Lab (Computing and Network Security Lab).
\endbio

\vspace{0.5in}

\bio{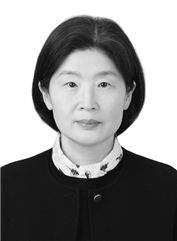}
\textbf{Kiyoung Kim} received the B.S. and M.S. degrees in Computer Science and Statistics from Chonnam National University in 1988, 1993, and Ph.D. degree in Computer Science from Chungbuk National University in Korea in 2002. Since 1988, she has been with the Electronics and Telecommunications Research Institute (ETRI), Daejeon, Korea, where she is currently a principal researcher. Her main research interests include network security, blockchain, and security and privacy for metaverse systems.
\endbio

\vspace{0.2in}

\bio{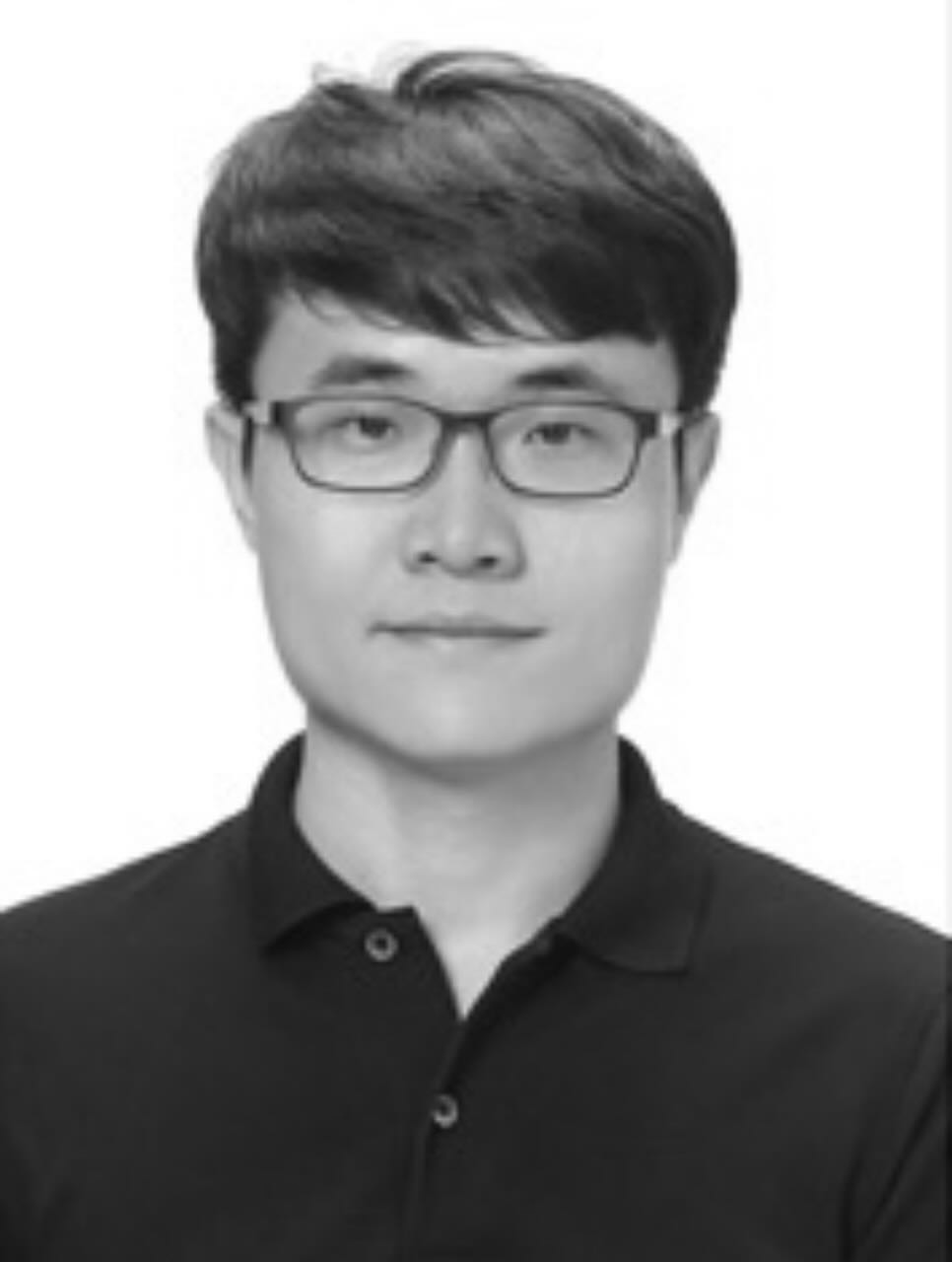}
\textbf{Hwanjo Heo} is a Principal Researcher at the Electronics and Telecommunications Research Institute (ETRI). He earned his Ph.D. from KAIST and his Master’s degree from Purdue University. His research interests include the security of computer systems, encompassing blockchain, computer networks, and artificial intelligence systems.
\endbio

\vspace{0.5in}

\bio{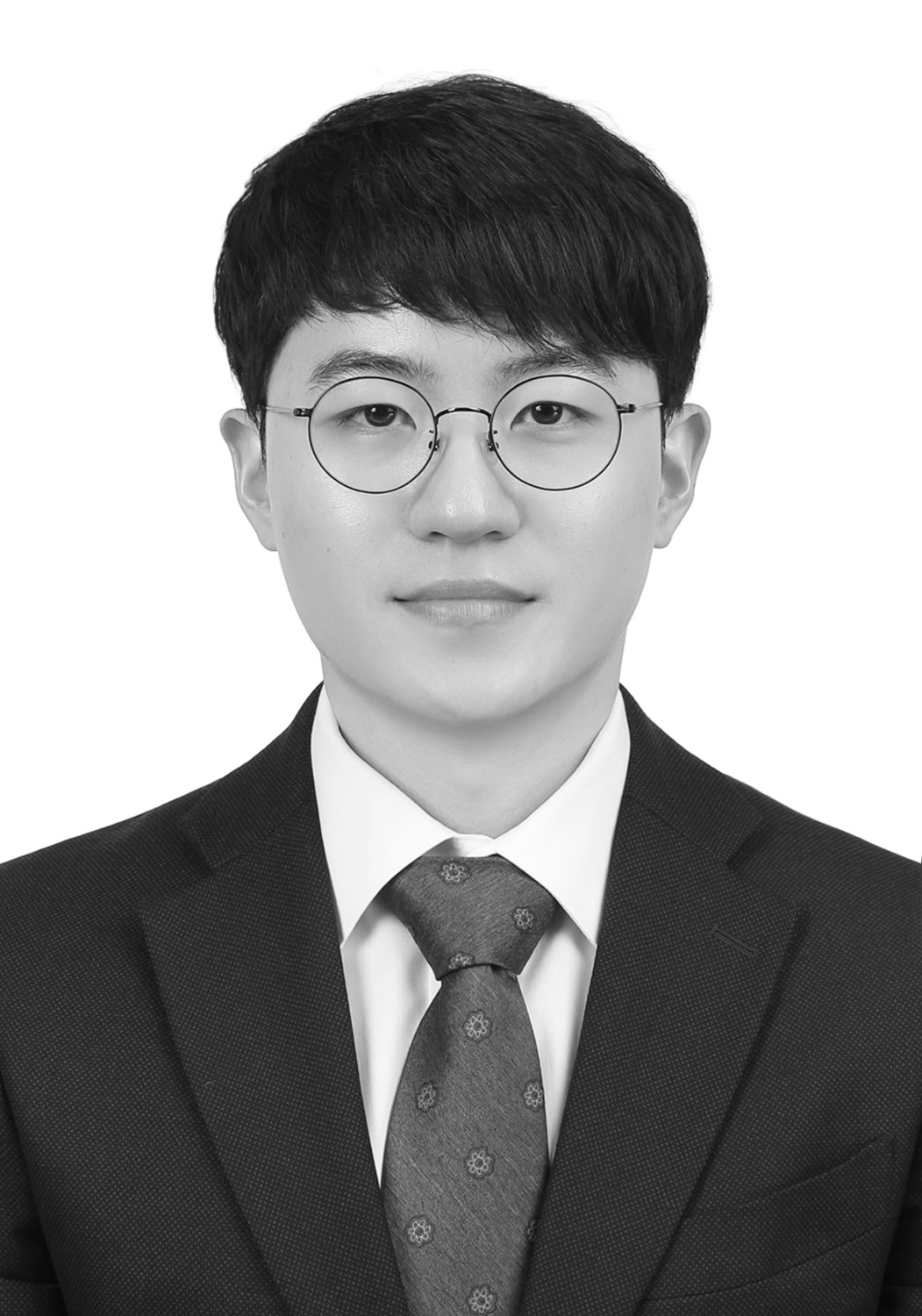}
\textbf{Jinwoo Kim} is an Assistant Professor in the Department of Computer Engineering at Chungbuk National University, Cheongju, Republic of Korea. He received his Ph.D. from the School of Electrical Engineering at KAIST, his M.S. degree from the Graduate School of Information Security at KAIST, and his B.S. degree from Chungnam National University in Computer Science and Engineering. He was previously an Assistant Professor in the School of Software at Kwangwoon University, Seoul, Republic of Korea. His research interests focus on investigating security and privacy issues in software-defined networks, cloud systems, and VR/XR.
\endbio

\end{document}